\documentclass[aps,prl,preprint,nopacs,superscriptaddress]{revtex4}
\usepackage{graphicx}
\usepackage{verbatim}
\usepackage{mathrsfs}
\usepackage{amsmath,amsfonts,amssymb}

\def\3{2.8in}    %used for figure widths
\def\2{2.5in}
\def\4{3.0in}

\def \beq {\begin{equation}}
\def \eeq {\end{equation}}
\begin{document}

\title{Hedgehog spin texture and Berry's phase tuning in a magnetic topological insulator}
\author{Su-Yang Xu}\affiliation {Joseph Henry Laboratory, Department of Physics, Princeton University, Princeton, New Jersey 08544, USA}
\author{Madhab Neupane}\affiliation {Joseph Henry Laboratory, Department of Physics, Princeton University, Princeton, New Jersey 08544, USA}
\author{Chang Liu}\affiliation {Joseph Henry Laboratory, Department of Physics, Princeton University, Princeton, New Jersey 08544, USA}
\author{Duming Zhang}\affiliation {Department of Physics, The Pennsylvania State University, University Park, Pennsylvania 16802-6300, USA}
\author{Anthony Richardella}\affiliation {Department of Physics, The Pennsylvania State University, University Park, Pennsylvania 16802-6300, USA}

\author{L. Andrew Wray}\affiliation {Joseph Henry Laboratory, Department of Physics, Princeton University, Princeton, New Jersey 08544, USA}\affiliation {Advanced Light Source, Lawrence Berkeley National Laboratory, Berkeley, California 94305, USA}

\author{Nasser Alidoust}\affiliation {Joseph Henry Laboratory, Department of Physics, Princeton University, Princeton, New Jersey 08544, USA}

\author{Mats Leandersson}\affiliation {MAX-lab, P.O. Box 118, S-22100 Lund, Sweden}
\author{Thiagarajan Balasubramanian}\affiliation {MAX-lab, P.O. Box 118, S-22100 Lund, Sweden}

\author{Jaime S\'anchez-Barriga}\affiliation {Helmholtz-Zentrum Berlin f\"ur Materialien und Energie, Elektronenspeicherring BESSY II, Albert-Einstein-Str. 15, D-12489 Berlin, Germany}
\author{Oliver Rader}\affiliation {Helmholtz-Zentrum Berlin f\"ur Materialien und Energie, Elektronenspeicherring BESSY II, Albert-Einstein-Str. 15, D-12489 Berlin, Germany}

\author{Gabriel Landolt}\affiliation {Swiss Light Source, Paul Scherrer Institute, CH-5232, Villigen, Switzerland}\affiliation {Physik-Institute, Universitat Zurich-Irchel, CH-8057 Zurich, Switzerland}
\author{Bartosz Slomski}\affiliation {Swiss Light Source, Paul Scherrer Institute, CH-5232, Villigen, Switzerland}\affiliation {Physik-Institute, Universitat Zurich-Irchel, CH-8057 Zurich, Switzerland}
\author{Jan Hugo Dil}\affiliation {Swiss Light Source, Paul Scherrer Institute, CH-5232, Villigen, Switzerland}\affiliation {Physik-Institute, Universitat Zurich-Irchel, CH-8057 Zurich, Switzerland}
\author{J\"urg Osterwalder}\affiliation {Physik-Institute, Universitat Zurich-Irchel, CH-8057 Zurich, Switzerland}

\author{Tay-Rong Chang}\affiliation {Department of Physics, National Tsing Hua University, Hsinchu 30013, Taiwan}
\author{Horng-Tay Jeng}\affiliation {Department of Physics, National Tsing Hua University, Hsinchu 30013, Taiwan}\affiliation {Institute of Physics, Academia Sinica, Taipei 11529, Taiwan}

\author{Hsin Lin}\affiliation {Department of Physics, Northeastern University, Boston, Massachusetts 02115, USA}
\author{Arun Bansil}\affiliation {Department of Physics, Northeastern University, Boston, Massachusetts 02115, USA}
\author{Nitin Samarth}\affiliation {Department of Physics, The Pennsylvania State University, University Park, Pennsylvania 16802-6300, USA}
\author{M. Zahid Hasan}\affiliation {Joseph Henry Laboratory, Department of Physics, Princeton University, Princeton, New Jersey 08544, USA}\affiliation{Princeton Center for Complex Materials, Princeton Institute for Science and Technology of Materials, Princeton University, Princeton, New Jersey 08544, USA}


\date{\today}
\maketitle
\textbf{
Understanding and control of spin degrees of freedom on the surfaces of topological materials are the key to future applications as well as for realizing novel physics such as the axion electrodynamics associated with time-reversal symmetry breaking on the surface. We experimentally demonstrate magnetically induced spin reorientation phenomena simultaneous with a Dirac-metal to gapped-insulator transition on the surfaces of manganese-doped Bi$_2$Se$_3$ thin films. The resulting electronic groundstate exhibits unique hedgehog-like spin textures at low energies which directly demonstrates the mechanics of time-reversal symmetry breaking on the surface. We further show that an insulating gap induced by quantum tunneling between surfaces exhibits spin texture modulation at low energies but respects time-reversal invariance. These spin phenomena and the control of their Fermi surface geometrical phase first demonstrated in our experiments pave the way for future realization of many predicted exotic magnetic phenomena of topological origin.}

\bigskip
\bigskip

Since the discovery of three dimensional topological insulators \cite{RMP, Moore, Kane PRB, David Nature BiSb, Qi PRB}, topological order proximity to ferromagnetism has been considered as one of the core interests of the field \cite{Cava Fe, Salman Fe, Yu Science QAH, Galvanic effect, Essin PRL, Zhang Axion, Hor PRB BiMnTe, Yayu WAL, Fe XMCD, Chen Science Fe, Andrew Nature physics Fe}. Such interest is strongly motivated by the proposed time-reversal (TR) breaking topological physics such as quantized anomalous chiral Hall current, spin current, axion electrodynamics, and inverse spin-galvanic effect \cite{Yu Science QAH, Galvanic effect, Essin PRL, Zhang Axion}, all of which critically rely on finding a way to break TR symmetry on the surface and utilize the unique TR broken spin texture for applications. Since quantum coherence is essential in many of these applications, devices need to be engineered into thin films in order to enhance or de-enhance surface-to-surface coupling or the quantum tunneling of the electrons. The experimental spin behavior of surface states under the two extreme limits, namely the doped magnetic groundstate and ultra-thin film quantum tunneling groundstate, is thus of central importance to the entire field. However, surprisingly, it is not known what happens to the spin configuration under these extreme conditions relevant for device fabrications. Fundamentally, TR symmetry is inherently connected to the Kramers' degeneracy theorem which states that when TR symmetry is preserved, the electronic states at the TR invariant momenta have to remain doubly spin degenerate. Therefore, the establishment of TR breaking effect fundamentally requires measurements of electronic groundstate with a \textit{spin}-sensitive probe. Here we utilize spin-resolved angle-resolved photoemission spectroscopy to measure the momentum space spin configurations in systematically magnetically doped, non-magnetically doped, and ultra-thin quantum coherent topological insulator films \cite{Xue Nature physics QL}, in order to understand the nature of electronic groundstates under two extreme limits vital for magnetic topological devices. These measurements allow us to make definitive conclusions regarding magnetism on topological surfaces, and make it possible to quantitatively isolate the TR breaking effect in generating the surface electronic gap from many other physical or chemical changes also leading to gap-like behavior \cite{Ando QPT, Haim Nature physics BiSe, vdW, Helical metal, Gap, Hofmann} often observed on the surfaces. Spin reorientation measurements and the systematic methodology demonstrated here can be utilized to probe quantum magnetism on the surfaces of other materials as well.

\bigskip
\bigskip
\textbf{Evolution of topological surface states with magnetic doping}
\newline
In order to study the evolution of topological surface states upon magnetic doping, magnetically (Mn\%) and (chemically similar) non-magnetically (Zn\%) doped Bi$_2$Se$_3$ thin films are prepared in high quality using the molecular beam epitaxy (MBE) growth method. A sample layout and a photograph image of a representative MBE grown film used for experiments are shown in Fig. 1a and b. Using standard X-ray magnetic circular dichroism \cite{XMCD, XMCD2}, we characterize the magnetic properties of doped Bi$_2$Se$_3$ films (Fig. 1c). In Mn-doped Bi$_2$Se$_3$, a hysteretic behavior in the out-of-plane magnetic response is observed consistently which suggests a ferromagnetically ordered groundstate. The observation of ferromagnetic character and its absence in Zn-Bi$_2$Se$_3$ motivate us to systematically compare and contrast the electronic density of state behavior in the vicinity of the surface Dirac node of these samples. Fig. 1d shows the measured electronic states of Mn(Zn)-doped Bi$_2$Se$_3$ using high-resolution (spin-integrated) angle-resolved photoemission spectroscopy (ARPES). In the undoped Bi$_2$Se$_3$ film (leftmost panel of Fig. 1d), a map of spectral density of states reveals a bright and intact Dirac node (signaled by the red spot located at the Dirac crossing in the panel), which indicates that in undoped Bi$_2$Se$_3$ the Dirac node is gapless, consistent with the previous studies of pure Bi$_2$Se$_3$ \cite{Matthew Nature physics BiSe}. In samples where Mn atoms are doped into the bulk (first row of Fig. 1d), we observe that the corresponding bright (red) spot at the Dirac node gradually disappears, revealing a clear systematic spectral weight suppression (SWS) with increasing Mn concentration. In contrast, the spectral intensity at the Dirac node is observed to survive upon systematic Zn doping except for the Zn = 10\% sample where some suppression of intensity is observed. This suggests that the Dirac node remains largely intact upon Zn doping. The observed systematic behavior of spectral evolution motivates us to quantitatively define an energy scale, $\textrm{E}_{\textrm{SWS}}$, associated with the SWS observed at the Dirac node. The $\textrm{E}_{\textrm{SWS}}$ is taken as the energy spacing between the upper Dirac band minimum and the Dirac node location along the energy axis as illustrated in Fig. 1e, which roughly corresponds to half of the surface gap magnitude. The value of the energy scale can be quantitatively determined by fitting the ARPES measured energy-momentum distribution curves (method described in supplementary section VI). The doping dependence of the $\textrm{E}_{\textrm{SWS}}$ on samples measured at $\textrm{T}=20$ K is shown in Fig. 2c. The $\textrm{E}_{\textrm{SWS}}$ is observed to increase rapidly with Mn concentration but it remains nearly zero with Zn doping. The effect of temperature dependence on $\textrm{E}_{\textrm{SWS}}$ is shown in Fig. 2d. The temperature induced decrease of $\textrm{E}_{\textrm{SWS}}$ is consistent with gradual weakening of magnetism. These observations collectively reveal direct correlation between X-ray magnetic circular dichroism measured ferromagnetic character and the ARPES measured SWS (or gap) on the Mn-Bi$_2$Se$_3$ films.

Although the predominant trend in the doping evolution of surface states suggests a correlation between magnetism and SWS, we do notice a non-negligible $\textrm{E}_{\textrm{SWS}}$ at high Zn concentration (10\%). This is likely due to increasing chemical disorder on the surface of the film since disorder degrades the surface quality. Similarly, magnetization measurements show that ferromagnetism vanishes before reaching $\textrm{T}=300$ K (inset of Fig. 2b), but nonzero SWS (Fig. 2d) is still observed perhaps, in this case, due to thermal disorder of the relaxed film surface at high temperatures. Therefore, the correlation between ferromagnetism and the ARPES gap is only clear at low temperatures in samples with reduced disorder. The momentum width ($\Delta{k}$) of the surface electronic states can be taken as a rough relative measure of surface disorder, sample to sample, which is found to significantly increase upon both magnetic and non-magnetic dopings (Fig. 2c). Strong spatial fluctuations of the surface electronic states in doped Bi$_2$Se$_3$ has been observed in a recent STM work by Beidenkopf \textit{et al.} \cite{Haim Nature physics BiSe}, where the authors suggest the observation of gap-like feature at the Dirac point without breaking TR symmetry. These ambiguities \cite{Gap} associated with the observed gap-like feature across many different experiments strongly call for critically important spin-resolved measurements which also serve as a collective method, as we show, to unambiguously identify the correct nature of the gap.

\bigskip
\textbf{Spin configuration of the magnetic groundstate}
\newline
In order to study the evolution of spin texture upon magnetic doping, we perform spin-resolved measurements on Mn-Bi$_2$Se$_3$ topological surface states. We present two independent but representative spin-resolved ARPES measurements on Mn(2.5\%)-Bi$_2$Se$_3$ film I and film II. Films I and II, both containing same nominal Mn concentration, are measured and analyzed using two different spectroscopic modes, namely, spin-resolved momentum distribution curve (spin-resolved MDC) measurement mode and spin-resolved energy distribution curve (spin-resolved EDC) measurement mode (see Methods section), in order to exclude any potential systematic error in the spin measurements. Fig. 3a-d show measurements on film I. Our data shows that out-of-plane spin polarization $\textrm{P}_z$ is nearly zero at large momentum $k_{//}$ far away from the Dirac point energy ($0<\textrm{E}_{\textrm{B}}<0.1$ eV in Fig. 3c, d).  While approaching the Dirac point ($0.1$ eV$<\textrm{E}_{\textrm{B}}<0.3$ eV), an imbalance between the spin-resolved intensity in $+\hat{z}$ and $-\hat{z}$ is observed (Fig. 3c). The imbalance is found to become systematically more pronounced in the data set where scans are taken by lowering the energy toward the Dirac point. This systematic behavior observed in the data reveals a net significant out-of-plane spin polarization in the vicinity of the ``gapped'' Dirac point or near the bottom of the surface state conduction band. More importantly, the out-of-plane spin component $\textrm{P}_z$ does not reverse its sign in traversing from $-k_{//}$ to $+k_{//}$. Such behavior is in sharp contrast to the spin textures observed in pure Bi$_2$Se$_3$ \cite{David Nature tunable} where spins point to opposite directions on opposite sides of the Fermi surface as expected from TR symmetry. Therefore, our $\textrm{P}_z$ measurements on film I near the gap edge reveal the TR breaking nature of the Mn-Bi$_2$Se$_3$ sample where magnetic hysteresis was observed using X-ray magnetic circular dichroism technique. In order to directly measure the spin of the surface state at $\bar{\Gamma}$ (the Kramers' momentum, $k_{//}=0$), we perform spin measurements on Mn-Bi$_2$Se$_3$ film II (same Mn concentration as film I) working in the spin-resolved EDC mode. The measured out-of-plane spin polarization ($\textrm{P}_z$) is shown in Fig. 3g. We focus on the $\textrm{P}_z$ measurement at $\bar{\Gamma}$, the Kramers' momentum $k_{//}=0$ (red curve): the surface electrons at TR invariant $\bar{\Gamma}$ are clearly observed to be spin polarized in the out-of-plane direction. The opposite sign of $\textrm{P}_z$ for the upper and lower Dirac band (red curve in Fig. 3g) shows that the Dirac point spin degeneracy is indeed lifted up ($\textrm{E}(k_{//}=0,\uparrow){\neq}\textrm{E}(k_{//}=0,\downarrow)$). Such observation directly counters the Kramers' theorem and therefore manifestly breaks the TR symmetry on the surface. Next we analyze $\textrm{P}_z$ measurements at finite $k_{//}$ (green curves in Fig. 3g) to extract the detailed configuration of the spin texture. In going to larger $k_{//}$ away from the $\bar{\Gamma}$ momenta, the measured $\textrm{P}_z$ is found to gradually decrease to zero. Moreover, the constant energy momentum space plane at the Dirac point ($\textrm{E}_\textrm{B}=\textrm{E}_\textrm{D}$) is observed to serve as a mirror plane that reflects all of the out-of-plane spin components between the upper and lower Dirac bands. Thus both spin-resolved MDC (film I) and spin-resolved EDC (film II) measurement modes result in the consistent conclusions regarding the spin configuration of the films. These systematic measurements, especially at the vicinity of the gap, reveal a hedgehog-like spin configuration for each upper (or lower) Dirac band separated by the magnetic gap, which breaks TR symmetry, as schematically presented in the inset of Fig. 3f (see supplementary for additional data and analysis).

\bigskip
\textbf{Spin configurations of the non-magnetic groundstates}
\newline
Spin texture measurements on non-magnetically doped films Zn(1.5\%)-Bi$_2$Se$_3$ are presented in Fig. 4a-d. The out-of-plane polarization $\textrm{P}_z$ measurements reveal a sharp contrast to the magnetically doped Mn-Bi$_2$Se$_3$ films, specifically, the near absence of finite $\textrm{P}_z$ component around $\bar{\Gamma}$ within our experimental resolution (Fig. 4d). A very small $\textrm{P}_z$, however, at large $k_{//}$ is observed, which is expected due to surface state warping also observed in other topological insulator (TI) compounds \cite{Liang Fu Warping} (Fig. 4d). The signal being associated with warping is further confirmed in our data due to their TR symmetric nature, that $\textrm{P}_z$ is observed to reverse its sign in traversing from $-k_{//}$ to $+k_{//}$. Moreover, our in-plane spin measurements (Fig. 4c) show that Zn-Bi$_2$Se$_3$ film retains the helical spin texture protected by the TR symmetry, as observed in pure Bi$_2$Se$_3$ and Bi$_2$Te$_3$ single crystals \cite{David Nature tunable}. Therefore we conclude that nonmagnetic Zn doping does not induce observable spin reorientation on the topological surface. The contrasting behavior observed between Mn-Bi$_2$Se$_3$ and Zn-Bi$_2$Se$_3$ samples as presented in Fig. 3 and 4 provide clear evidence for TR symmetry breaking in Mn-Bi$_2$Se$_3$. 

A surface band gap at the Dirac point can also be generated in Bi$_2$Se$_3$ in its ultra-thin film limit. In this case, the top and bottom surfaces couple together and open up a gap as electrons can tunnel from one to the other (Fig. 4e-h). Such a gap in the surfaces is not related to magnetism. It is important to know the spin configuration associated with such a gap. In Fig. 4e-h, we utilize Spin-resolved ARPES to measure the spin configuration on the very top region (within 5 $\textrm{\AA}$) of a Bi$_2$Se$_3$ film whose thickness is three quintuple layers (3 QL $\simeq28.6$ $\textrm{\AA}$). At large parallel-momenta far away from $\bar{\Gamma}$ (e.g. -0.10 $\textrm{\AA}^{-1}$ in Fig. 4g), we observe clear spin polarization following left-handed helical configuration with the magnitude of the polarization around $35-40$\%. However, in going to smaller $k_{//}$, the magnitude of the spin polarization is observed to be reduced. At the TR invariant $\bar{\Gamma}$ momenta, spin-resolved measurements (Fig. 4g red curve) show no net spin polarization. This reduction of the spin polarization at small momenta near the gap is an intrinsic effect (see supplementary Fig. S19). These observations can be understood by considering the scenario where the surface-to-surface coupling dominates, and the two energetically degenerate surface states from top and bottom that possess opposite helicities of the spin texture cancel each other at $\bar{\Gamma}$ \cite{Xue Nature physics QL}. This results in strong suppression of spin polarization in the vicinity of this gap, whereas upon probing momenta to large $k_{//}$ away from $\bar{\Gamma}$, the finite kinetic energy of the surface states (${\propto}$ $vk_{//}$) naturally leads to the spatial decoupling of two Dirac cones. These spin measurements on the ultra-thin Bi$_2$Se$_3$ film reveal the interplay between quantum tunneling (coupling) and the spin texture modification, which is of importance in spin-based device design with thin films. The observed spin texture however does not break TR symmetry, since the spins remain doubly degenerate at the TR invariant momenta $\bar{\Gamma}$. This is in clear contrast to the spin texture observed in Mn-Bi$_2$Se$_3$.

\bigskip
\textbf{Magnetic contribution and geometrical phase tuning}
\newline
The magnetic contribution to the gap of the Mn-Bi$_2$Se$_3$ film can be quantitatively identified using the spin texture data. The simplest $\mathbf{k}{\cdot}\mathbf{p}$ Hamiltonian that describes topological surface states under TR symmetry breaking can be written as $\textrm{H}=v(k_{x}\sigma_{y}-k_{y}\sigma_{x})+b_z\sigma_z$, where $\sigma$ and $k$ are the spin and momentum operators respectively, $b_z$ corresponds to half of the magnetic gap and $v$ is the velocity of the surface Dirac band. We specify the out-of-plane polar angle $\theta$ of the spin polarization vector (inset of Fig. 3g) as $\theta=arctan\frac{\textrm{P}_{z}}{\textrm{P}_{//}}$. The magnitude of the polar angle $\theta$ reflects the competition between the out-of-plane TR breaking texture (${\propto}$ $b_z$) and the in-plane helical configuration component (${\propto}$ $vk_{//}$). Using the measured spin-resolved data sets $(\theta, k)$, we fit the magnetic interaction strength $b_z$ within a $\mathbf{k}{\cdot}\mathbf{p}$ scenario (see supplementary section II.2). As an example, we fit $b_z$ based on spin-resolved data sets in Fig. 3g on Mn(2.5\%)-Bi$_2$Se$_3$ film II, as shown in Fig. 5c and obtain a value of 21 meV. This is a significant fraction of the SWS energy scale observed on the same sample, $\textrm{E}_{\textrm{SWS}}>50$ meV (see Fig. 2c for Mn = 2.5\%) obtained from the spin-integrated measurements in Fig. 1c. Thus we identify the magnitude of the magnetic contribution ($b_z$) to the observed spectral weight suppression using spin-sensitive measurements, which suggests that the magnetic contribution is significant to $\textrm{E}_{\textrm{SWS}}$ in Mn-Bi$_2$Se$_3$. 

As demonstrated recently \cite{Hasan QPT}, the geometrical phase (GP) defined on the spin texture of the surface state Fermi surface \cite{Kane PRB} (also known as the Berry's phase) bears a direct correspondence to the bulk topological invariant realized in the bulk electronic band structure via electronic band inversion \cite{David Nature tunable, Hasan QPT}. We experimentally show that a GP tunability can be realized on our magnetic films which is important to prepare the sample condition to the axion electrodynamics limit. On the Mn-Bi$_2$Se$_3$ film, spin configuration pattern can be understood as a competition between the out-of-plane TR breaking component and the in-plane helical component of spin. The in-plane spin that can be thought of winding around the Fermi surface in a helical pattern contributes to a nonzero GP \cite{David Nature tunable}, whereas the out-of-plane TR breaking spin direction is constant as one loops around the Fermi surface hence does not contribute to the Berry's phase (GP). As a result, the GP remains almost $\pi$ if the chemical potential lies at energies far away from the Dirac point, whereas it starts to decrease and eventually reach to 0 as one approaches the TR breaking gap by lowering the chemical potential as discussed in theory \cite{Shen WAL}, at least within the magnetic energy scale $b_z$ (Fig. 5). We show that this theoretical requirement can be experimentally achieved on the Mn-Bi$_2$Se$_3$ surface via surface NO$_2$ adsorption at some specific dosage level. Fig. 5a shows the Mn(2.5\%)-Bi$_2$Se$_3$ surface states with \textit{in situ} NO$_2$ adsorption. The chemical potential is observed to be gradually shifted and finally placed within the magnetic gap. The associated phase (GP) at each experimentally achieved sample chemical potential is found to gradually change from $\pi$ to $0$. The GP $= 0$ is the experimental condition for realizing axion electrodynamics with a topological insulator \cite{Essin PRL, Zhang Axion}.

With the chemical potential moved into the magnetic gap, the time-reversal breaking in-gap state features a singular hedgehog-like spin texture (Fig. 5d). Such spin configuration simultaneous with the chemical potential placed within the magnetic gap (Fig. 5d) is the fundamental requirement for most of the theoretical proposals relevant to the utilization of magnetic topological insulators in novel devices \cite{Qi PRB, Yu Science QAH, Galvanic effect, Essin PRL, Zhang Axion}. Furthermore, if the bulk band-gap of the Mn-Bi$_2$Se$_3$ film is tuned to zero at the critical point of the topological phase transition \cite{Hasan QPT}, a new topologically protected Weyl semimetal phase \cite{Gil, Weyl Balents} with yet more exotic but undiscovered state of matter is also predicted to take place, which is among the most exciting future frontiers to be enabled by our achievement of a sample that features a TR broken hedgehog-like spin texture with a GP = 0 state \cite{Gil, Weyl Balents}.

\bigskip
\bigskip
\textbf{Methods}
\newline
\textbf{Sample growth:} The Mn doped Bi$_2$Se$_3$ thin films were synthesized by Molecular Beam Epitaxy using high purity elemental (5N) Mn, Bi, and Se sources. A thin GaAs buffer layer was first deposited on the epi-ready GaAs 111A substrate after thermal desorption of the native oxide under As pressure. Then the substrate was transferred to another chamber without breaking the vacuum, where a second buffer layer of ZnSe was deposited for further smoothing the surface. Mn-Bi$_2$Se$_3$ layer ($\sim60$ nm) was then grown with a high Se/Bi beam equivalent pressure (BEP) ratio of $\sim15$. The Mn doping concentration was controlled by adjusting the Bi/Mn BEP ratio ranging from 8 to 60. To protect the surface from oxidation, a thick Se capping layer was deposited on the Mn-Bi$_2$Se$_3$ thin film immediately after the growth. The Zn doped Bi$_2$Se$_3$ control samples were also synthesized under the same conditions as Mn-Bi$_2$Se$_3$, with the Zn doping concentration controlled by Bi/Zn BEP. Ultra-thin Bi$_2$Se$_3$ film was prepared in thickness of 3 QL, with a typical 1 QL peak to peak variation ($2-4$ QL). Complete details of film growth are presented in Ref \cite{Duming} and also detailed in the supplementary.

\textbf{Electronic structure measurements:} Angle-resolved photoemission spectroscopy (ARPES) measurements (spin-integrated) were performed with $29$ eV to $64$ eV photon energy on beamlines 10.0.1 and 12.0.1 at the Advanced Light Source (ALS) in Lawrence Berkeley National Laboratory (LBNL). Spin-resolved ARPES measurements were performed on the I3 beamline at Maxlab in Lund, Sweden, COPHEE spectrometer SIS beamline at the Swiss Light Source (SLS) in Switzerland, and UE112-PGM1 beamline PHOENEXS chamber at BessyII in Berlin, Germany, using photon energies of $8-11$ eV, $20-22$ eV, and $55-60$ eV for the three beamlines respectively. For ARPES measurements, samples were annealed \textit{in situ} to evaporate the amorphous Se cap layer in order to reveal the clean surface (see supplementary for further details). 

\textbf{Spin-resolved measurements:} Spin-resolved measurements were performed using spin-resolved ARPES beamlines mentioned above with double classical Mott detectors and linearly polarized photons. The spin-resolved measurements in Fig. 3 and 4 were performed in two different modes: spin-resolved MDC mode and spin-resolved EDC mode. spin-resolved MDC mode means each single measurement measures the spin polarization at a fixed binding energy ($\textrm{E}_\textrm{B}$) along certain momentum cut direction in momentum space (see Fig. 3a-d on Mn(2.5\%)-Bi$_2$Se$_3$ film I). spin-resolved EDC mode means each single measurement measures the spin polarization at a fix momentum ($k$) along the binding energy axis ($\textrm{E}_\textrm{B}$) (see Fig. 4e-g and Fig. 4 on Mn(2.5\%)-Bi$_2$Se$_3$ film II, Zn(1.5\%)-Bi$_2$Se$_3$ film, and 3 QL undoped Bi$_2$Se$_3$ film). The details of the spin-resolved technique and principles can be found in supplementary.

\textbf{Magnetic property characterizations:} In order to cross-check the ferromagnetism implied by the spin measurements on Mn-Bi$_2$Se$_3$ surface, X-ray magnetic circular dichroism (XMCD) was independently performed on Mn-Bi$_2$Se$_3$ film surface at the back-endstation of D1011 beamline at Maxlab in Lund, Sweden, with total electron yield (TEY) mode at temperatures ranging from 40 K to 300 K. XMCD measurements were performed on the $\textrm{L}_{23}$ absorption edge of the Mn atom by the standard methods \cite{XMCD}, which is widely used to study the magnetic properties of transition metals and dilute magnetic semiconductor thin films or monolayers \cite{XMCD, XMCD2}.

\bigskip
The detailed principles and methods of all techniques used in the experiments including sample growth, ARPES and spin-resolved ARPES, NO$_2$ surface adsorption, and XMCD are further elaborated in the supplementary information.

\bigskip
\bigskip
\textbf{Acknowledgements}
\newline
Work at Princeton University is supported by the US National Science Foundation Grant, NSF-DMR-1006492. M.Z.H. acknowledges visiting-scientist support from Lawrence Berkeley National Laboratory and additional support from the A. P. Sloan Foundation. The spin-resolved and spin-integrated photoemission measurements using synchrotron X-ray facilities are supported by The Swedish Research Council ,The Knut and Alice Wallenberg Foundation, the Swiss Light Source, the Swiss National Science Foundation, the German Federal Ministry of Education and Research, and the Basic Energy Sciences of the US Department of Energy. Theoretical computations are supported by the US Department of Energy (DE-FG02-07ER46352 and AC03-76SF00098) as well as the National Science Council and Academia Sinica in Taiwan, and benefited from the allocation of supercomputer time at NERSC and Northeastern University's Advanced Scientific Computation Center. Sample growth and characterization are supported by US DARPA (N66001-11-1-4110). We gratefully acknowledge Alexei Preobrajenski for beamline assistance on XMCD measurements at D1011 beamline at Maxlab in Lund Sweden. We also thank Sung-Kwan Mo and Alexei Fedorov for beamline assistance on spin-integrated photoemission measurements in Lawrence Berkeley National Laboratory.

\bigskip
\bigskip
\textbf{Author contributions}
\newline
S.-Y.X. performed the experiments with assistance from M.N., C.L., L.A.W., N.A., and M.Z.H.; D.M.Z, A.R., and N.S. provided samples; M.L., T.B., J.S.B., O.R., G.L., B.S., J.H.D., and J.O. provided beamline assistance; T.-R.C., H.-T.J., H.L., and A.B. carried out the theoretical calculations; M.Z.H. was responsible for the overall direction, planning and integration among different research units.

\bigskip
\bigskip

\*Correspondence should be addressed to M.Z.H. (Email: mzhasan@princeton.edu).

\newpage
\begin{figure*}
\centering
\includegraphics[width=17cm]{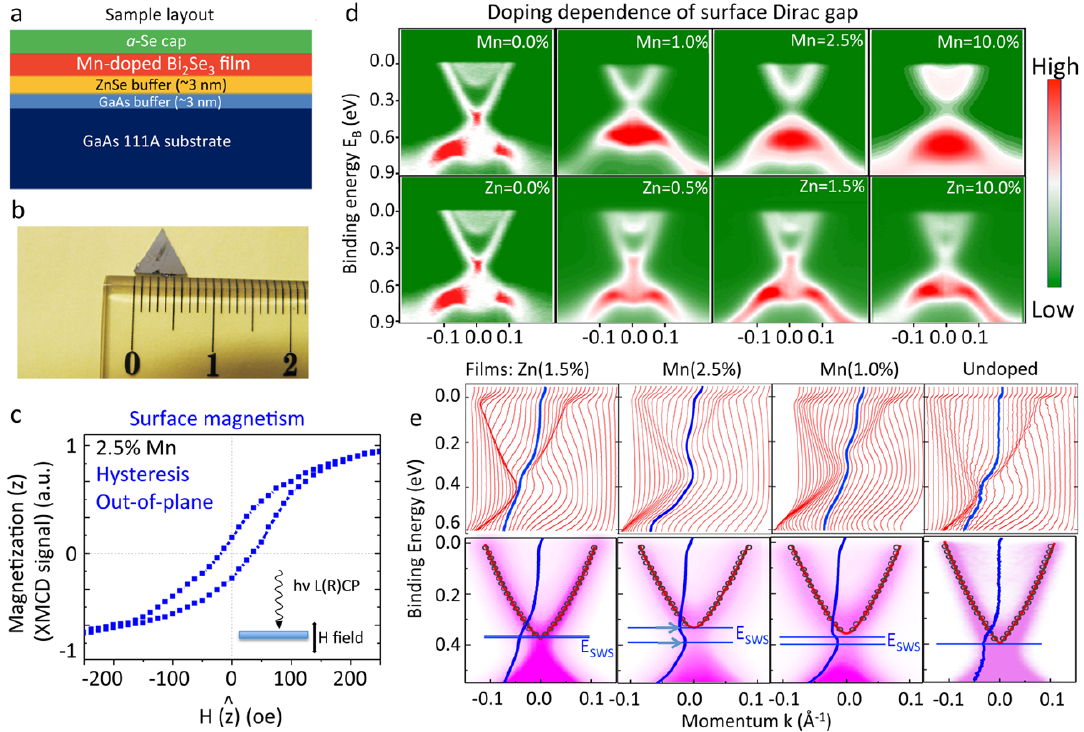}
\caption{\textbf{Magnetic (Mn\%) and non-magnetic (Zn\%) doping on Bi$_2$Se$_3$ films. a and b,} Schematic layout and photograph of MBE grown Bi$_2$Se$_3$ films used for the experiments.  \textbf{c,} Magnetization measurement at $\textrm{T}=45$ K using magnetic circular dichroism shows out-of-plane ferromagnetic character of the Mn-Bi$_2$Se$_3$ film (111) surface through the observed hysteretic response (detailed in supplementary section II.5.2). Inset shows the measurement geometry. L(R)CP represents left(right)-handed circularly polarized light. \textbf{d,} Electronic band dispersion of Mn(Zn)-doped Bi$_2$Se$_3$ MBE thin films along the $\bar{\textrm{M}}-\bar{\Gamma}-\bar{\textrm{M}}$ momentum space cut. The doping level (noted in the top-right corner of each panel) reflects nominal Mn(Zn) concentration, which is defined as the ratio of $\frac{\textrm{Mn(Zn)}}{\textrm{Mn(Zn)}+\textrm{Bi}}$ over the entire film crystal. Increasing Mn concentration leads to spectral weigh suppression at the Dirac point, signaling a ``gap''. The ``gap'' and the hysteresis are possibly correlated in the data. \textbf{e,} Energy-momentum distribution curves of Mn(Zn)-doped Bi$_2$Se$_3$ samples. The energy scale associated with the spectral weight suppression (SWS) $\textrm{E}_{\textrm{SWS}}$ is observed as the energy spacing between the upper Dirac band minimum and the Dirac point location along the energy axis.} 
\end{figure*}

\begin{figure*}
%\centering
\includegraphics[width=15cm]{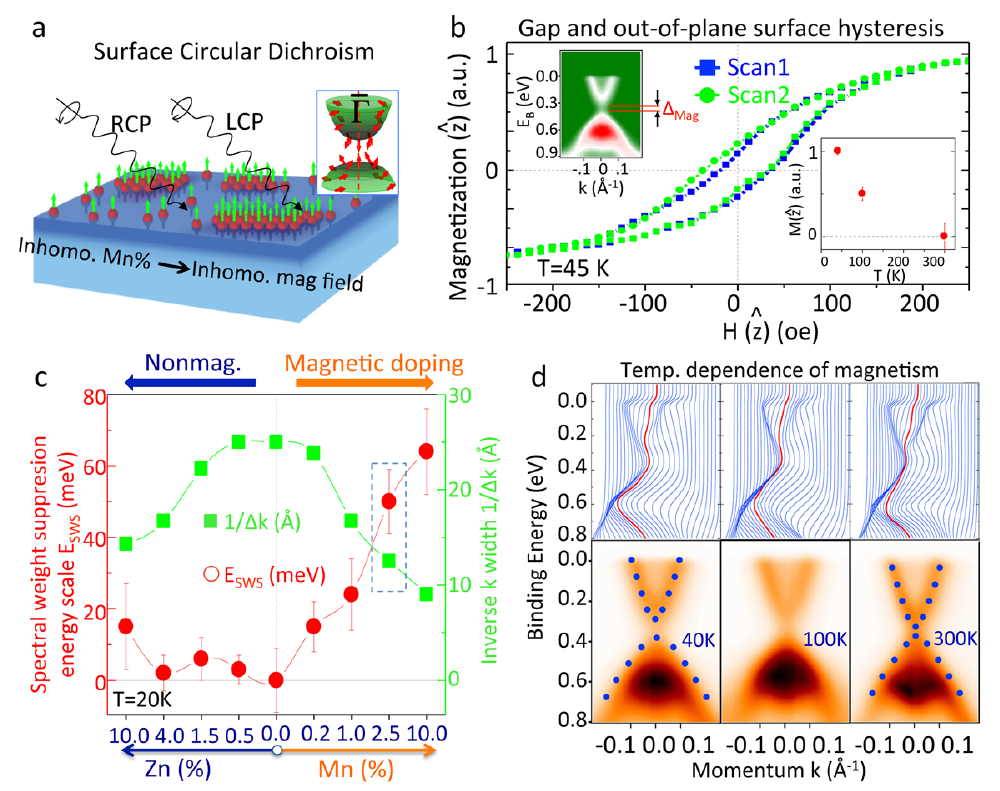}
\caption{\textbf{Temperature and doping dependence of magnetically induced changes on Mn-Bi$_2$Se$_3$ surface. a,} The Mn atoms on the surface of the film are out-of-plane magnetically ordered, serving as a local magnetic field which results in the spin texture (see Fig. 3) reorientation. \textbf{b,} Two independent hysteresis measurements at $\textrm{T}=45$ K using X-ray magnetic circular dichroism reveal the ferromagnetic character of the Mn-Bi$_2$Se$_3$ film surface. The lower inset shows the remanent surface magnetization as a function of temperature. The out-of-plane magnetic hysteresis and ARPES gap were found to be correlated with each other. The upper inset shows the gap at the Dirac point in Mn(2.5\%)-Bi$_2$Se$_3$ film. \textbf{c,} The spectral weight suppression energy scale $\textrm{E}_{\textrm{SWS}}$ (defined in Fig. 1) and inverse momentum width $1/{\Delta}k$ of the surface states are shown as a function of Mn and Zn concentration measured at $\textrm{T}=20$ K. \textbf{d,} Temperature dependence of spectral weight suppression energy scale around the Dirac point of Mn(2.5\%)-Bi$_2$Se$_3$ film (as noted in panel \textbf{c} by the dotted square). The $\textrm{E}_{\textrm{SWS}}$ decreases as temperature is raised signaling gradual weakening of magnetism.}
\end{figure*}

\newpage
\begin{figure}[t]
\centering
\includegraphics[width=16cm]{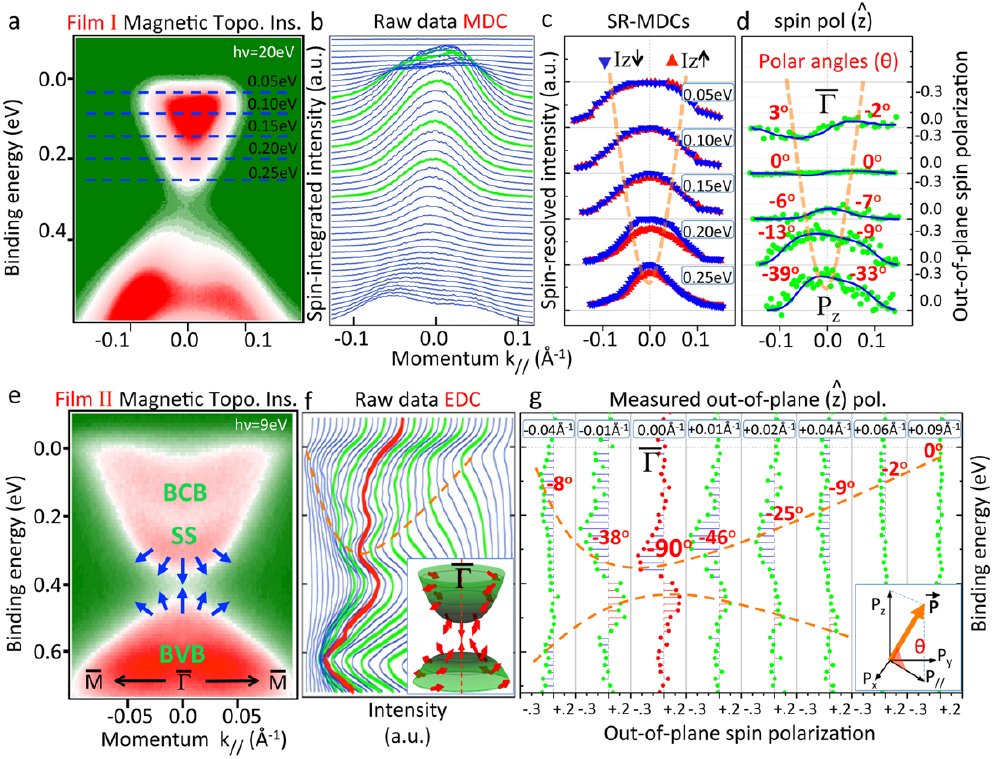}
\caption{\textbf{Spin configuration measurements of a magnetic topological insulator.} Panels \textbf{a-d} present spin-resolved measurements on film I using 20 eV photons in the momentum distribution curve (MDC) mode. Panels \textbf{e-g} present measurements on film II using 9 eV photons in the energy distribution curve (EDC) mode. \textbf{a and b,} Spin-integrated data and corresponding MDCs. \textbf{c,} Spin-resolved MDC spectra for out-of-plane direction as a function of electron binding energy. \textbf{d,} Measured out-of-plane component of the spin polarization, presented in terms of respective out-of-plane polar angles ($\theta$) defined in the inset of \textbf{g}. \textbf{e-f,} Spin-integrated dispersion and EDCs. The EDCs selected for detailed spin-resolved measurements are highlighted in green and red. \textbf{g,} Measured out-of-plane spin polarization of the EDCs corresponding to panel \textbf{f}. Inset defines the definition of spin polarization vector $\vec{\mathbf{\textrm{P}}}$ and the polar angle $\theta$. The momentum value of each spin-resolved EDC is noted on the top. The polar angles ($\theta$) of the spin vectors obtained from measurements are also noted. The $90^{\circ}$ polar angle observed at $\bar{\Gamma}$ suggests that the spin vector at $\bar{\Gamma}$ momenta points in the vertical direction. The spin behavior at $\bar{\Gamma}$ and its surrounding momentum space reveals a hedgehog-like spin configuration for each Dirac band separated by the gap, which breaks time-reversal symmetry ($\textrm{E}(\vec{k}=0,\uparrow){\neq}\textrm{E}(\vec{k}=0,\downarrow)$), as schematically shown in the inset of \textbf{f}.}
\end{figure}

\newpage
\begin{figure*}
%\centering
\includegraphics[width=17cm]{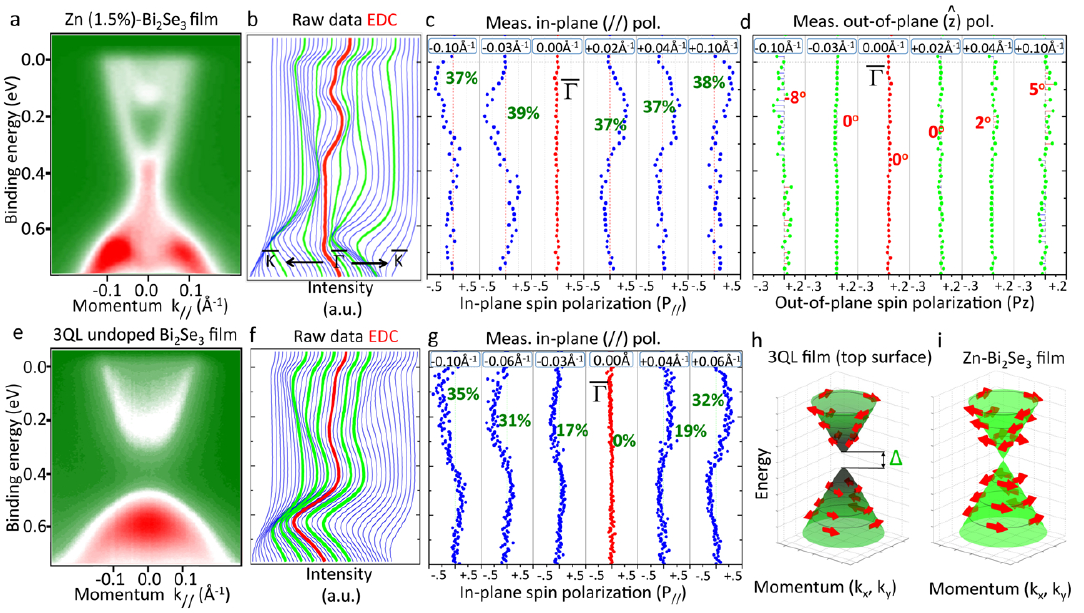}
\caption{\textbf{Spin configurations on non-magnetic samples. a-d,} Spin-resolved measurements on 1.5\% non-magnetic Zn-Bi$_2$Se$_3$ film. The in-plane polarization measurements \textbf{(c)} reveal the helical spin configuration, as in pure Bi$_2$Se$_3$ topological insulator \cite{David Nature tunable}, suggesting that non-magnetic impurities do not induce spin reorientation on the topological surfaces. Out-of-plane measurements \textbf{(d)} show that no significant out-of-plane spin polarization $\textrm{P}_z$ is induced near the $\bar{\Gamma}$ point (the time-reversal invariant momenta), leaving the system time-reversal invariant overall. \textbf{e-g,} Spin-resolved ARPES measurements on ultra-thin undoped Bi$_2$Se$_3$ film of three quintuple-layer thickness. The net spin polarization is found to be significantly reduced near the gap edge around the $\bar{\Gamma}$ momenta. This is consistent with the fact that in ultra-thin films electrons tunnel between the top and bottom surfaces. \textbf{h and i,} A schematic of the two types of spin textures observed in our data. }
\end{figure*}

\begin{figure*}
%\centering
\includegraphics[width=17cm]{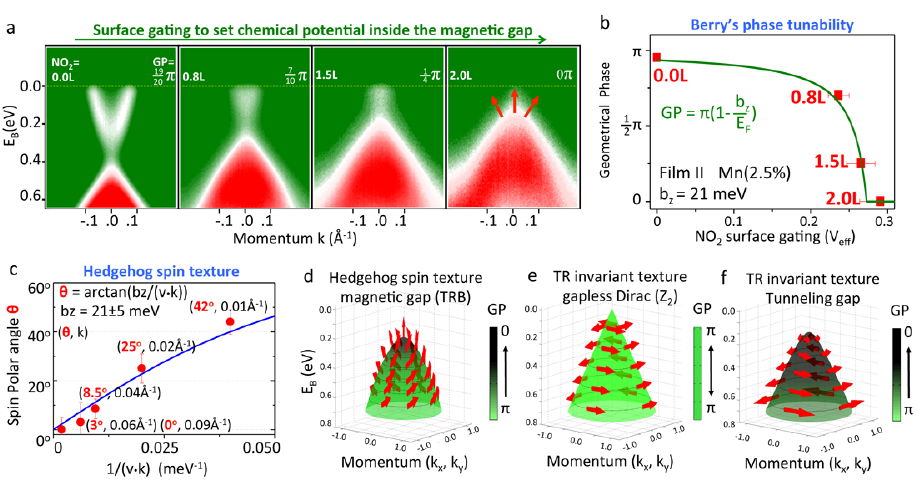}
\caption{\textbf{Chemical potential tuned to lie inside the magnetic gap. a,} Measured surface state dispersion upon \textit{in situ} NO$_2$ surface adsorption on the Mn-Bi$_2$Se$_3$ surface. The NO$_2$ dosage in the unit of Langmuir ($1\textrm{L}=1\times10^{-6}$ torr${\cdot}$sec) and the tunable geometrical phase (see text) associated with the topological surface state are noted on the top-left and top-right corners of the panels, respectively. The red arrows depict the time-reversal breaking out-of-plane spin texture at the gap edge based on the experimental data. \textbf{b,} Geometrical phase (GP) associated with the spin texture on the iso-energetic contours on the Mn-Bi$_2$Se$_3$ surface as a function of effective gating voltage induced by NO$_2$ surface adsorption. Red squares represent the GP experimentally realized by NO$_2$ surface adsorption, as shown in \textbf{a}. GP $= 0$ (NO$_2$ = $2.0$ L) is the condition for axion dynamics. \textbf{c,} The magnetic interaction strength $b_z$ (see text for definition), which corresponds to half of the magnetic gap magnitude, is obtained based on spin-resolved data sets (polar angle $\theta$, momentum k) for Mn(2.5\%)-Bi$_2$Se$_3$ film II (see Fig. 3e-g). \textbf{d,} The time-reversal breaking spin texture features a singular hedgehog-like configuration when the chemical potential is tuned to lie within the magnetic gap, corresponding to the experimental condition presented in the last panel in \textbf{a}.  \textbf{e and f,} Spin texture schematic based on measurements of Zn-doped Bi$_2$Se$_3$ film (60 QL), and 3 QL undoped ultra-thin film with chemical potential tuned onto the Dirac point energy or within the tunneling gap.} 
\end{figure*}

\clearpage
\setcounter{figure}{0}
\renewcommand{\figurename}{\textbf{Supplementary Figure}}

\textbf{
\begin{center}
{\large \underline{Supplementary lnformation}: \\Hedgehog spin texture and Berry's phase tuning in a magnetic topological insulator}
\end{center}
}

\vspace{0.2cm}

\begin{center}
Su-Yang Xu, M. Neupane, Chang Liu, D. M. Zhang, A. Richardella, L. A. Wray,\\
N. Alidoust, M. Leandersson, T. Balasubramanian,  J. S\'anchez-Barriga,\\
O. Rader, G. Landolt, B. Slomski, J. H. Dil, J. Osterwalder, T.-R. Chang, \\
H.-T. Jeng, H. Lin, A. Bansil, N. Samarth, and M. Z. Hasan
\end{center}

\vspace{0.25cm}

\begin{figure*}[h]
\centering
\includegraphics[width=17cm]{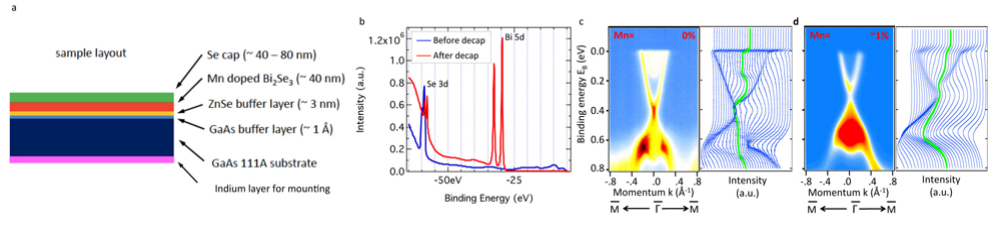}
\caption{\label{Decap} \textbf{Demonstration of sample surface preparation (decapping) procedure in ARPES measurements.} \textbf{a,} Sample layout of the MBE grown Mn(Zn)-doped Bi$_2$Se$_3$ films. \textbf{b,} Core level spectroscopies on MBE thin film before and after the decapping procedure. \textbf{c-d,} typical spin-integrated ARPES dispersion mapping and corresponding energy dispersion curves (EDCs) of undoped and $\sim1\%$ Mn-doped (nominal concentration) Bi$_2$Se$_3$ thin film along high symmetry $\bar{\textrm{M}}-\bar{\Gamma}-\bar{\textrm{M}}$ direction.}
\end{figure*}

\textbf{
\begin{center}
{\large This file includes:\\}
\end{center}
}
\vspace{0.45cm}
\textbf{
\begin{tabular}{l l}
\underline{SI I.} & Materials and methods \\
\underline{SI II.} & Mn-Bi$_2$Se$_3$ (Spin-resolved measurements, first principles electronic structure\\
 & calculation, and magnetic property characaterzation) \\ 
\underline{SI III.} & Zn-Bi$_2$Se$_3$ (Spin-resolved measurements)\\
\underline{SI IV.} & Ultra-thin undoped Bi$_2$Se$_3$ (Spin-resolved measurements)\\
\end{tabular}
}
\\\\
\ \textbf{FigS\ref{Decap} to S\ref{QL_Resolution}} \\

\newpage

\textbf{{\large SI I. Materials and methods}}

\bigskip

Spin-integrated angle-resolved photoemission spectroscopy (ARPES) measurements were performed with $29$ eV to $64$ eV photon energy on beamlines 10.01 and 12.01 at the Advanced Light Source (ALS) in Lawrence Berkeley National Laboratory (LBNL). Spin-resolved ARPES measurements were performed on the I3 beamline at Maxlab \cite{I3 1,I3 2} in Lund, Sweden, COPHEE spectrometer SIS beamline at the Swiss Light Source (SLS) in Switzerland \cite{Hoesch Spin ARPES, Hugo PRB}, and UE112-PGM1 beamline PHOENEXS chamber at BessyII in Berlin, Germany, using the classical Mott detectors and photon energies of $8-11$ eV, $20-22$ eV, and $55-60$ eV for the three beamlines respectively. For ARPES measurements, samples were decapped \textit{in situ} to prepare the clean surface (see below), and measured at temperature below $50$ K at chamber pressure less than $2 {\times} 10^{-10}$ torr at all beamlines. 

The Mn doped Bi$_2$Se$_3$ thin films were synthesized by Molecular Beam Epitaxy (MBE) using high purity elemental (5N) Mn, Bi, and Se sources. A thin GaAs buffer layer was first deposited on the epi-ready GaAs 111A substrate after thermal desorption of the native oxide under As pressure. Then the substrate was transferred to another chamber without breaking the vacuum, where a second buffer layer of ZnSe was deposited to further smooth the surface. Mn-Bi$_2$Se$_3$ layer ($\sim60$ nm) was then grown with a high Se/Bi beam equivalent pressure (BEP) ratio of $\sim15$. The Mn doping concentration was controlled by adjusting the Bi/Mn BEP ratio ranging from 8 to 60. To protect the surface from oxidation, a thick Se capping layer was deposited on the Mn-Bi$_2$Se$_3$ thin film immediately after the growth (see Fig. ~\ref{Decap}a). The Zn doped Bi$_2$Se$_3$ control samples were also synthesized under the same conditions as Mn-Bi$_2$Se$_3$, with the Zn doping concentration controlled by Bi/Zn BEP. Ultra-thin Bi$_2$Se$_3$ film was prepared in thickness of 3 QL, with a typical 1 QL peak to peak variation ($2\sim4$ QL). 

In order to reveal the clean Bi$_2$Se$_3$ surface needed for photoemission measurements, the MBE-grown thin films are heated up inside the ARPES chamber to $\sim250^{\circ}C$ under vacuum better than $1\times10^{-9}$ torr to blow off the Se capping layer on top of the Bi$_2$Se$_3$ film. Fig.~\ref{Decap}b demonstrates the decapping process by ARPES core level spectroscopies measurement. Only selenium core level is observed before the decapping process (blue curve), whereas both selenium and bismuth peaks are shown after the decapping (red curve). ARPES measurements are then performed on the clean thin film surface. Fig.~\ref{Decap}c-d show a typical spin-integrated ARPES measured dispersion mappings after decapping of both undoped Bi$_2$Se$_3$ and $\sim1\%$ (nominal concentration, same as the maintext) Mn-doped Bi$_2$Se$_3$ thin film. 

SR-ARPES measurements are performed with double Mott detector setup, which enables to measure spin-resolved photoemission spectra along momentum $k_{//}$ at fixed binding energy $\textrm{E}_\textrm{B}$ (SR-MDC) or along binding energy $\textrm{E}_\textrm{B}$ at fixed momentum $k_{//}$ (SR-EDC) for all three spatial directions (I$\hat{x}{\uparrow}{\downarrow}$, $I\hat{y}{\uparrow}{\downarrow}$, and $I\hat{z}{\uparrow}{\downarrow}$). The detailed and comprehensive review of SR-ARPES technique is presented in Ref. \cite{Hugo Review}. The spin-resolved data on Mn-Bi$_2$Se$_3$ shown in the main paper is done in remanent magnetization mode: the film is pre-magnetized in the UHV preparation chamber in the ARPES system at the base temperature by approaching the film with a permanent magnet with magnetic field roughly at 0.4T, in order to induce the remanent magnetization. The film is then transferred from the preparation chamber to the measurement chamber to perform spin-resolved measurements, maintaining the pressure and temperature.

The surface sensitivity of ARPES measurements can be obtained from the photoemission electron escape depth. The electron escape depth with incident photon energy from ($50-60$ eV) are well below 5 $\textrm{\AA}$. This is important to explain the SR-ARPES data on the ultra-thin 3 QL Bi$_2$Se$_3$ in the maintext, because a $<5$ $\textrm{\AA}$ surface sensitivity means SR-ARPES probes the very top quintuple layer of the 3 QL film, whose total thickness is around 3 nm.

Magnetization of the surface in the presence of Mn atoms are checked with X-ray magnetic circular dichroism (XMCD) using total electron yield (TEY) mode that greatly enhances the surface sensitivity (within 50 $\textrm{\AA}$) of the magnetization measurements. XMCD measures the dichroism signal of the $\textrm{L}_{23}$ absorption edge spectra, which is widely used to study the magnetic properties of transition metals and dilute magnetic semiconductor thin films or monolayers. The X-ray magnetic circular dichroism (XMCD) measurements are performed at the back-endstation of D1011 beamline at Maxlab in Lund, Sweden with total electron yield (TEY) mode on Mn-Bi$_2$Se$_3$ film. The temperature of the XMCD measurements ranges from 40 K to 300 K. The magnetic field is applied in the out-of-plane direction normal to the sample surface, with a maximum H field of 300 oe. The left (right) hand circularly polarized incident light is also normal to the film surface, with an accuracy of the circular polarization better than 90\%. The sample is prepared the same way as in ARPES measurements \textit{in situ} by decapping inside the XMCD UHV chamber. The XMCD technique and the XMCD measurements on Mn-Bi$_2$Se$_3$ film surface are discussed in detail in the section below SI. II. 3. 

Adsorption of NO$_2$ molecules on sample surface was achieved via controlled exposures to NO$_2$ gas (Matheson, 99.5\%). The adsorption effects were studied under static flow mode by exposing the clean sample surface to the gas for a certain time at the pressure of $1 \times 10^{-8}$ torr, then taking data after the chamber was pumped down to the base pressure. Spectra of the NO$_2$ adsorbed surfaces were taken within minutes of opening the photon shutter to minimize potential photon induced charge transfer and desorption effects.

The first-principles calculations are based on the generalized gradient approximation (GGA) \cite{GGA 1} using the full-potential projected augmented wave method \cite{GGA 2} as implemented in the VASP package \cite{GGA 3}. A $2\times2\times6$ quintuple-layer slab model with a vacuum thickness lager than 10 $\textrm{\AA}$ is used in this work. To simulate the dilute Mn doping, one Bi atom is substituted by one Mn atom in both the top and bottom layers with the internal atomic coordinates optimized. The electronic structure calculations were performed over $5\times5\times1$ Monkhorst-Pack k-mesh with the spin-orbit coupling included self-consistently.

\newpage
\textbf{{\large SI II. Mn-Bi$_2$Se$_3$}}
\begin{enumerate}

\item \textbf{Comprenhensive spin-resolved measurements}
\begin{figure*}[h]
\centering
\includegraphics[width=16cm]{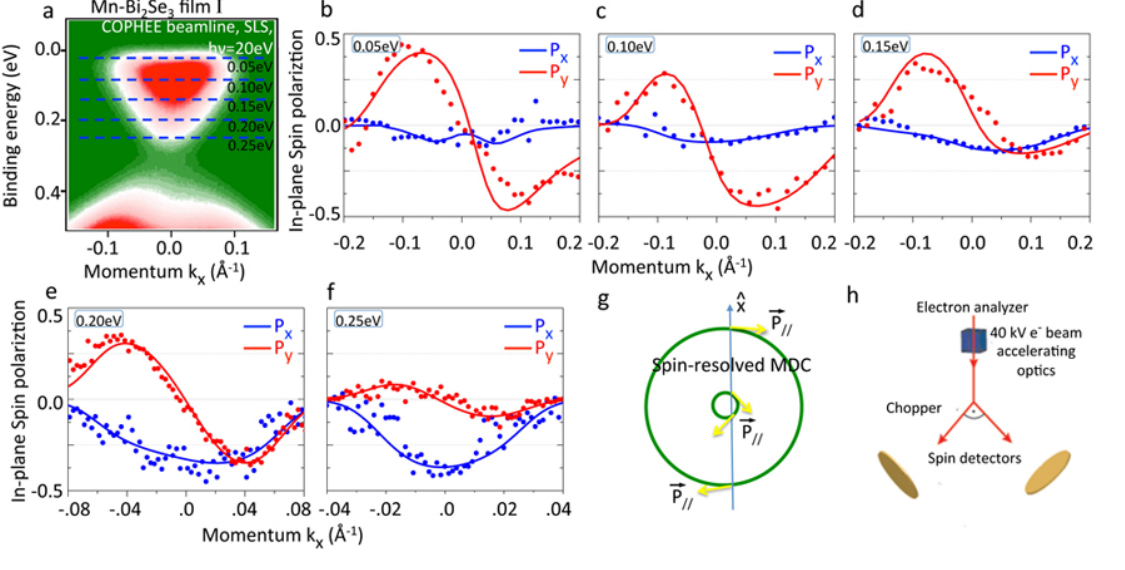}
\caption{\label{COPHEE_spin} \textbf{In-plane spin measurements of Mn-Bi$_2$Se$_3$ film I.} \textbf{a,} Spin-integrated ARPES dispersion mapping at incident photon energy 20 eV (used for spin-resolved measurements). \textbf{b-f,} In-plane spin polarization along $\hat{x}$ and $\hat{y}$ directions. \textbf{g,} The $\hat{x}$ direction is defined as along the spin-resolved MDC measurement direction.}
\end{figure*}

Here we show comprehensive spin-resolved data on Mn-Bi$_2$Se$_3$ film I and II in the maintext Fig. 3. Mn-Bi$_2$Se$_3$ film I is measured using the COPHEE spectrometer SIS beamline at the Swiss Light Source (SLS) in Switzerland (Fig.~\ref{COPHEE_spin}h) \cite{Hoesch Spin ARPES, Hugo PRB}. The measurements are performed in spin-resolved momentum distribution curve (MDC) mode (along momentum, with fixed energies, as shown in Fig.~\ref{COPHEE_spin}a). Since the out-of-plane component is already presented in the maintext (Fig. 3a-d), here we show the in-plane components of the spin-resolved measurements.

The $\hat{x}$ direction is defined as along the spin-resolved MDC measurement direction. Therefore, if the SR-MDC measurements perfectly cross the Brillouin zone (BZ) center $\bar{\Gamma}$ point, then polarization along $\hat{x}$ direction $\textrm{P}_x$ should be strictly zero, whereas $\textrm{P}_y$ should be opposite at the opposite side of the Fermi surface representing the helical spin texture \cite{David Nature tunable}. Fig.~\ref{COPHEE_spin}b-f show the measured in-plane spin polarization along both $\hat{x}$ and $\hat{y}$ directions. Indeed, $\textrm{P}_y$ reverses its sign when going from $-k$ to $+k$. Meanwhile, a finite $\textrm{P}_x$ is also observed which becomes increasingly significant when going to large binding energies close to the Dirac point. Such nonzero $\textrm{P}_x$  is owning to a small tilt of the sample, which results in a ``miss-cut'' away from the BZ center $\bar{\Gamma}$ point (see Fig.~\ref{COPHEE_spin}g). The observed $\textrm{P}_y$ (opposite sign at opposite sides of the Fermi surface) and $\textrm{P}_x$ (same sign at opposite sides of the Fermi surface) proves that the in-plane components of the spin texture still preserve the helical spin configuration.

\begin{figure*}[h]
\centering
\includegraphics[width=17cm]{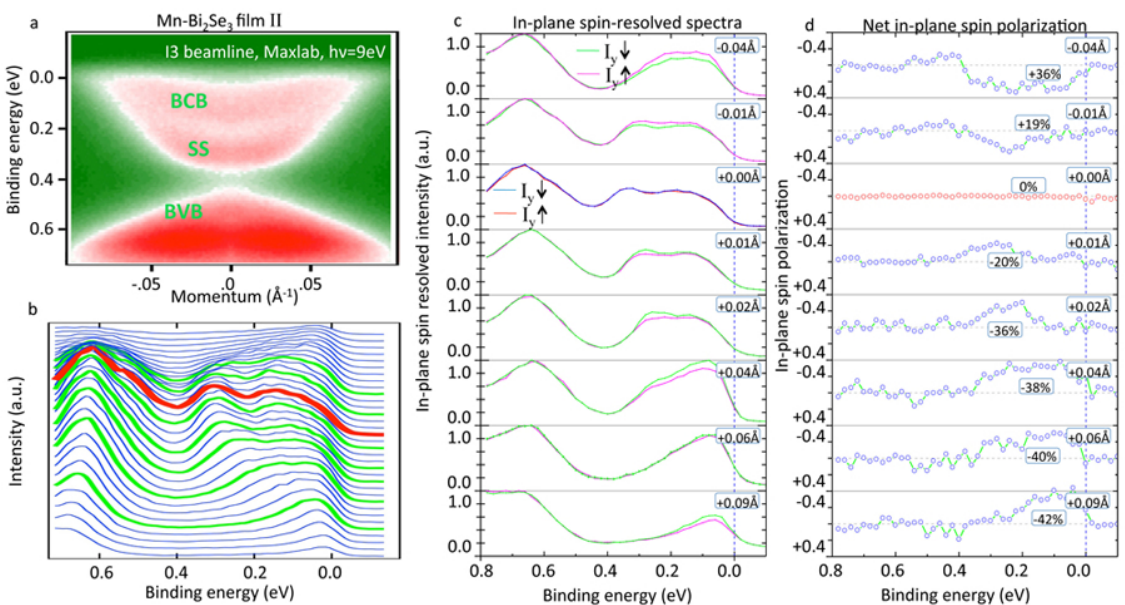}
\caption{\label{MAX_in_plane} \textbf{In-plane spin measurements of Mn-Bi$_2$Se$_3$ film II.} \textbf{a-b,} Spin-integrated ARPES measured dispersion and EDCs with incident photon energy 9 eV (used for spin-resolved measurements). The EDCs selected for spin-resolved measurements are highlighted in green (red) colors in the EDC panel. The EDC at the $\bar{\Gamma}$ momenta is in red color. \textbf{c,} In-plane spin-resolved EDC spectra. \textbf{d,} In-plane spin polarization obtained from \textbf{c}.}
\end{figure*}

\begin{figure*}[h]
\centering
\includegraphics[width=17cm]{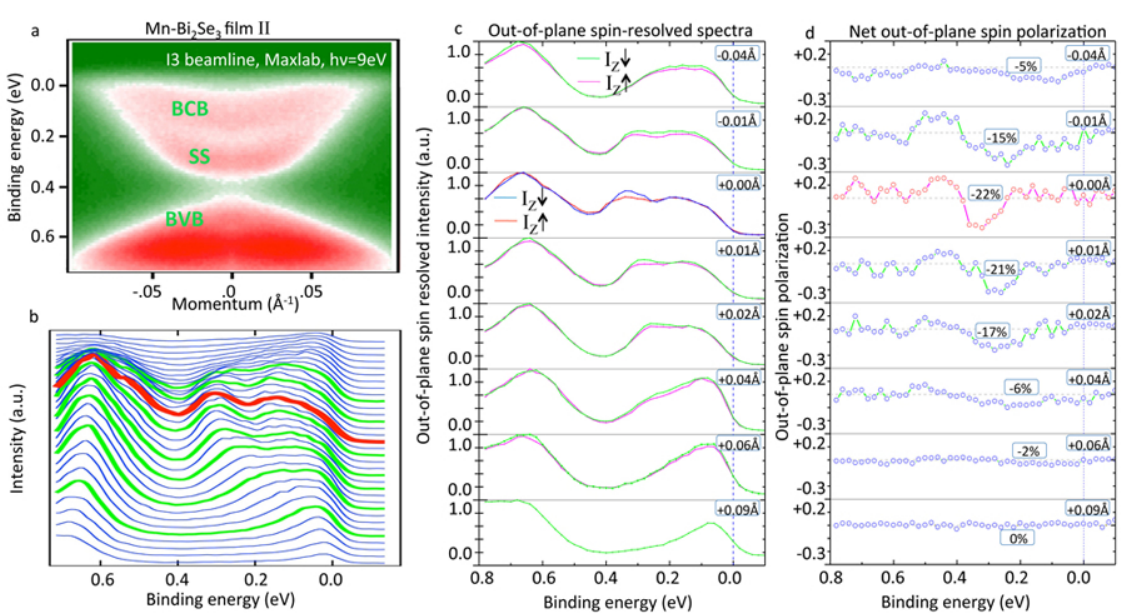}
\caption{\label{MAX_out_of_plane} \textbf{Out-of-plane spin measurements of Mn-Bi$_2$Se$_3$ film II.} \textbf{In-plane spin measurements of Mn-Bi$_2$Se$_3$ film II} \textbf{a-b,} Spin-integrated ARPES measured dispersion and EDCs with incident photon energy 9 eV (used for spin-resolved measurements). The EDCs selected for spin-resolved measurements are highlighted in green (red) colors in the EDC panel. The EDC at the $\bar{\Gamma}$ momenta is in red color. \textbf{c,} Out-of-plane spin-resolved EDC spectra. \textbf{d,} Out-of-plane spin polarization obtained from \textbf{c}.}
\end{figure*}

Mn-Bi$_2$Se$_3$ film II is measured at the I3 beamline at Maxlab \cite{I3 1,I3 2} in Lund, Sweden, with the spin-resolved energy distribution curve (EDC) mode (along binding energy, with fixed momenta). The in-plane and out-of-plane measurements are shown in Fig.~\ref{MAX_in_plane} and Fig.~\ref{MAX_out_of_plane} respectively. Since manipulator of the I3 beamline is equipped with an additional motor to tilt the sample, the small tilt due to sample mounting is corrected before measurements. As a result $\textrm{P}_x$ is zero in the measurements. Again, $\textrm{P}_y$ shows the ``classical'' helical spin texture. Moreover, $\textrm{P}_z$ shows the singly degenerate Dirac point and the TR breaking evidence, which has been presented in the maintext already.

\item \textbf{Surface magnetic interaction strength $b_z$ based on spin-resolved measurements on Mn-Bi$_2$Se$_3$}
\begin{figure*}[h]
\centering
\includegraphics[width=12cm]{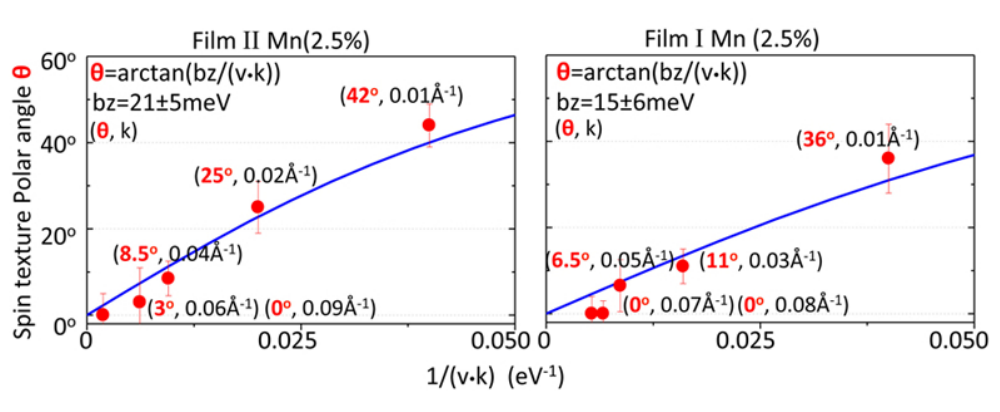}
\caption{\label{SR_fitting} \textbf{Surface magnetic interaction strength $b_z$ based on spin-resolved measurements.} The magnetic interaction strength can be fitted based on the spin-resolved measurements on films II and I (see maintext Fig. 3) of Mn(2.5\%)-Bi$_2$Se$_3$.}
\end{figure*}

Here we show the details of the method of fitting the magnetic interaction strength $b_z$ using the spin-resolved data sets on Mn-Bi$_2$Se$_3$, as presented in maintext Fig. 5c.

We start from the simplest $k{\cdot}p$ hamiltonian for the topological surface states under TR breaking: 
\begin{equation}
\textrm{H}=v(k_x\sigma_y-k_y\sigma_x)+b_z\sigma_z+o(k^2)
=\left( \begin{array}{ccc}
b_z & vk_+ \\
vk_- & -b_z \\
\end{array} \right)+o(k^2)
\end{equation}

where $b_z$ is the magnetic interaction strength, $k_+$ and $k_-$ are defined as $k_+=k_y+ik_x$ and $k_-=k_y-ik_x$ respectively. The higher $k$ orders $o(k^2)$ correspond to the parabolic bending of the linear dispersion at large momentum ($k^2$ term) and Fermi surface warping also at large momentum ($k^3$ term), which are not significant for surface states at the energy near the Dirac point. This description is in particularly appropriate for Bi$_2$Se$_3$ since the surface Dirac cone is nearly isotropic and the warping effect is not strong as comparing to Bi$_2$Te$_3$. We solve the Schr\"odinger equation 

\begin{equation}
\textrm{H}|\psi>=\textrm{E}|\psi>
\end{equation}

The eigen-values of the energy spectrum are 

\begin{equation}
\textrm{E}_{\pm}={\pm}\sqrt{(vk)^2+(b_z)^2}
\end{equation} 

which represent the upper ($\textrm{E}_+$) and the lower ($\textrm{E}_-$) Dirac bands respectively. The eigen-states wavefunction for upper ($|\psi_+>$) and lower ($|\psi_->$) Dirac bands are 

\begin{equation}
|\psi_+>=\frac{1}{\sqrt{2}}e^{i{\phi}/2}cos(\theta/2)|\frac{1}{2}>+e^{-i{\phi}/2}sin(\theta/2)|-\frac{1}{2}>
\end{equation}

\begin{equation}
|\psi_->=\frac{1}{\sqrt{2}}e^{i{\phi}/2}cos(\theta/2)|\frac{1}{2}>-e^{-i{\phi}/2}sin(\theta/2)|-\frac{1}{2}>
\end{equation}

where $|{\pm}\frac{1}{2}>$ are the eigen-states of electron spin in $\hat{z}$ direction, and $(\theta, \phi)$ is parameterized as 

\begin{equation}
(vk_x, vk_y, b_z)=\sqrt{(vk)^2+(b_z)^2}(cos(\theta)cos(\phi), cos(\theta)sin(\phi), sin(\theta))
\end{equation}

The angles $(\theta, \phi)$ exactly depict the spin texture, which are measured by spin-resolved ARPES. Where $\theta$ is the out-of-plane polar angle of the spin vector, and $\phi$ is the in-plane azimuthal angle of the spin vector. From equation (S6), we obtain that 
\begin{equation}
\theta=arctan(\frac{b_z}{vk})
\end{equation}
 
This equation (S7) reflects the competition between TR-breaking effect and spin helical texture. We can use equation (S7) to fit the magnetic interaction strength $b_z$ as presented in Fig. 5c of maintext.

As shown in Fig.~\ref{SR_fitting}, we have the fitting of $b_z$ using spin-resolved date sets from films II and I of Mn(2.5\%)-Bi$_2$Se$_3$ respectively. The spin-resolved data on films I and II of Mn(2.5\%)-Bi$_2$Se$_3$ are systematically presented in Fig. 3 of maintext, Fig.~\ref{COPHEE_spin}, Fig.~\ref{MAX_in_plane}, and Fig.~\ref{MAX_out_of_plane} in the supplementary. For each measured spin vector, both in-plane spin polarization $\textrm{P}_{//}$ and out-of-plane spin polarization $\textrm{P}_{z}$ are measured. The spin vector out-of-plane polar angle $\theta$ is then experimentally obtained as $tan{\theta}=\frac{\textrm{P}_{z}}{\textrm{P}_{//}}$. Thus for each experimentally measured spin vector, the data can be presented in the way of $(\theta, k)$, where $\theta$ is the experimentally measured out-of-plane polar angle $\theta$, and $k$ is the momentum location of that spin vector in the momentum space. Using these data sets, the magnetic interaction strength $b_z$ can be fitted by equation (S7). 

Fig.~\ref{SR_fitting} shows the fitting based on spin-resolved date sets from films II and I of Mn(2.5\%)-Bi$_2$Se$_3$ respectively. The fitted $b_z$ value is found to be $21\pm5$ meV for data sets fitting from film II, and $15\pm6$ meV for data sets fitting from film I.

\item \textbf{Remanent magnetization and temperature dependence of the TR breaking out-of-plane spin texture}

\begin{figure*}[h]
\centering
\includegraphics[width=10cm]{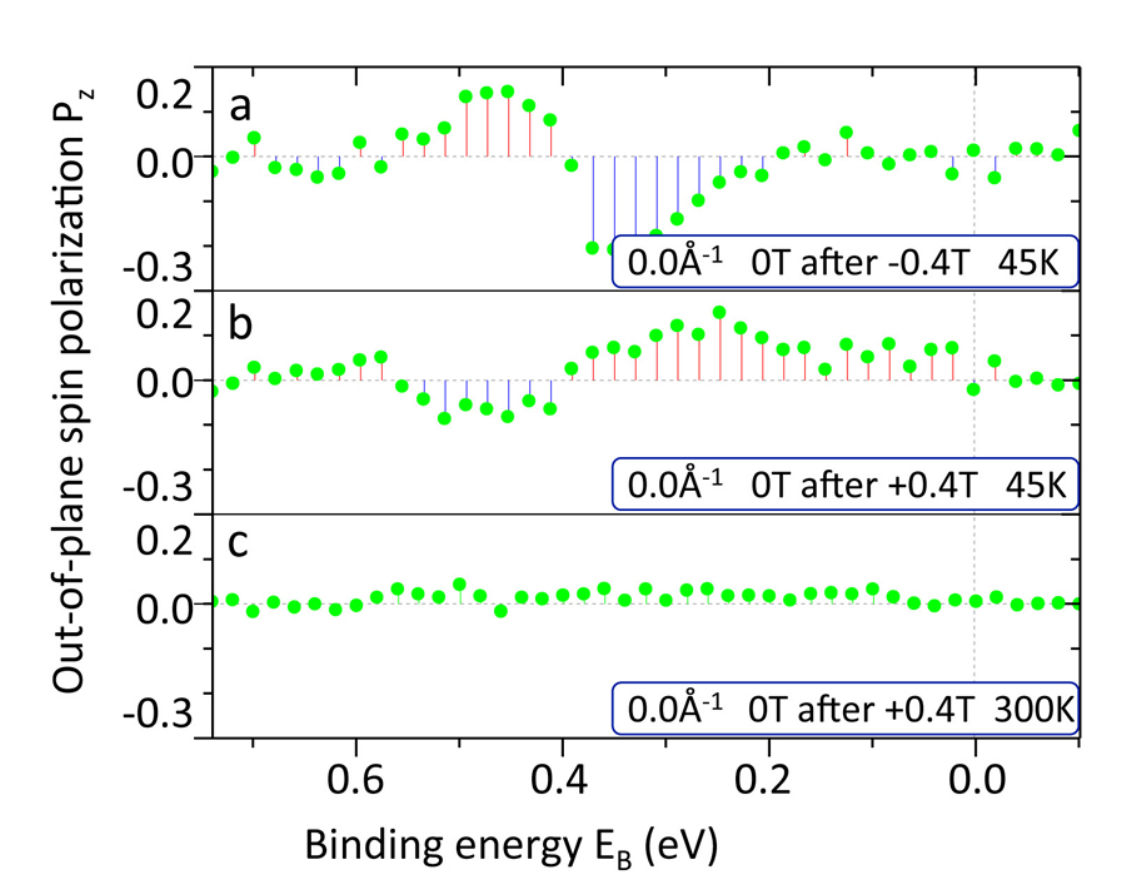}
\caption{\label{Field_dep} \textbf{Remanent magnetization and temperature dependence of the out-of-plane spin polarization of Mn-Bi$_2$Se$_3$ film surface.} \textbf{a,} Out-of-plane spin polarization $\textrm{P}_z$ with film magnetization in -z direction at T=45 K. Such condition is achieved by apply 0.4T magnetic field in -z direction in the prep chamber and then transfer the film into the measurement chamber. \textbf{b,} Out-of-plane spin polarization $\textrm{P}_z$ with film magnetization in +z direction at T=45 K. Such condition is achieved by apply 0.4T magnetic field in +z direction in the prep chamber and then transfer the film into the measurement chamber. \textbf{c,} Out-of-plane spin polarization $\textrm{P}_z$ at T=300 K. No magnetization is induced because temperature is above Curie temperature of the film surface (see below ferromagnetic order section).}
\end{figure*}

The spin-resolved data on Mn-Bi$_2$Se$_3$ shown in the main paper is done in remanent magnetization mode: the film is pre-magnetized in the UHV preparation chamber in the ARPES system at the base temperature by approaching the film with a permanent magnet with magnetic field roughly at 0.4T, in order to induce the remanent magnetization. The the film is transferred from the preparation chamber to the measurement chamber to perform spin-resolved measurements, maintaining the pressure and temperature.
 
Here we show the spin resolved measurements on  Mn-Bi$_2$Se$_3$ film II as a function of remanent magnetization directions and temperatures. The helical in-plane component of the spin polarization does not show any dependence with magnetization or temperature. Thus we focus on the out-of-plane component which is a direct consequence of the magnetization of the film surface. To capture the essence, we only show the out-of-plane spin polarization measurements at the TR breaking invariant $\bar{\Gamma}$ momenta $k=$0.0 $\textrm{\AA}^{-1}$. Fig.~\ref{Field_dep}a repeats the measurements shown in the maintext and in Fig.~\ref{MAX_out_of_plane}d. When the remanent magnetization is reversed, as shown in Fig.~\ref{Field_dep}b, the out-of-plane spin polarization is also observed to be reversed. Identical measurements at room temperature 300 K  (Fig.~\ref{Field_dep}c) show no observable out-of-plane spin polarization.

\item \textbf{First principles electronic structure calculation of Mn-Bi$_2$Se$_3$ and Mn impurity band}
\begin{figure*}[h]
\centering
\begin{center}
\includegraphics[width=13cm]{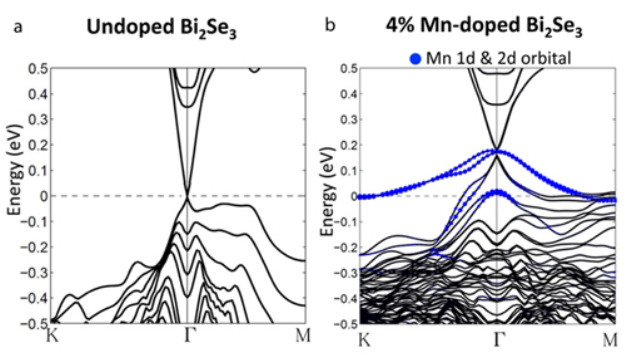}
\end{center}
\caption{\label{Mn_impurity} \textbf{First principles electronic structures calculation of undoped and Mn-doped Bi$_2$Se$_3$.} \textbf{a,} undoped Bi$_2$Se$_3$. \textbf{b} 4\% Mn-doped Bi$_2$Se$_3$. The band character of Mn 1d and 2d orbital are plotted on the band structure calculation.}
\end{figure*}
 
Here we show our first principles electronic structure calculation of the Mn-doped Bi$_2$Se$_3$. Fig.~\ref{Mn_impurity} shows calculated band structure of both undoped and 4\% Mn-doped Bi$_2$Se$_3$. Comparing to the undoped material, Mn doping is found to hole-dope the system, which moves the Fermi level (energy=0) below the Dirac point. The hole doping is due to the $2^+$ ionic state for Mn atom (as compared to a $3^+$ ionic state for the Bi atoms in Bi$_2$Se$_3$). Such p-type doing behavior is also directly observed in our ARPES measurements. 

Apart from moving the chemical potential, an important feature in the calculation is the development of extra electronic states with strong Mn orbital character (Mn impurity bands). The impurity bands are observed to be majorly located at energies lower than 0.3 eV below the Dirac poin in our calculation (see Fig.~\ref{Mn_impurity}b). At the same time, the valence band spectral weight is found to be shifted from $k_{//}{\simeq}\pm0.1$ $\textrm{\AA}^{-1}$ to $k_{//}{\simeq}0.0$ $\textrm{\AA}^{-1}$ by Mn doping, as shown in Fig. 1c in the maintext, measured with the same photon energy. The newly developed strong intensity at the valence band of Mn-Bi$_2$Se$_3$ in ARPES measurements can have strong correlations to the Mn impurity bands in the calculation.

\item \textbf{Ferromagnetic ordering in Mn-Bi$_2$Se$_3$ film}

\textbf{5.1. SQUID measurements of Mn-Bi$_2$Se$_3$ film}
\begin{figure*}[h]
\centering
\begin{center}
\includegraphics[width=17cm]{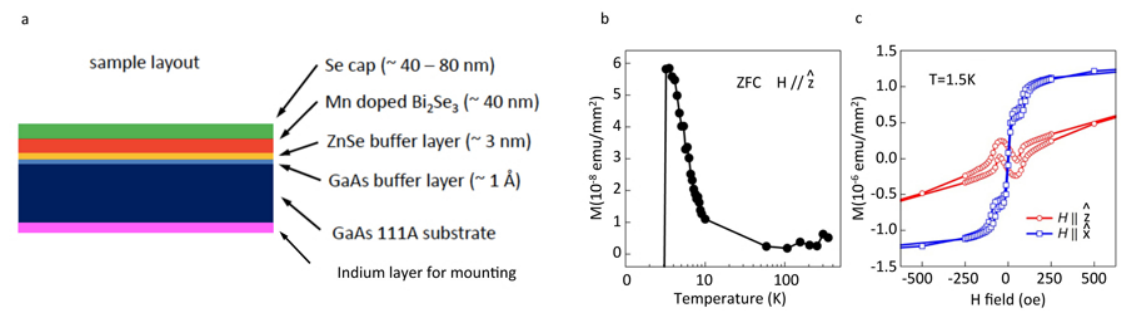}
\end{center}
\caption{\label{SQUID} \textbf{Magnetic property of Mn-Bi$_2$Se$_3$ measured by SQUID.} \textbf{a,} Mn-Bi$_2$Se$_3$ film sample layout. The indium layer on the bottom which is necessary for the sample mounting for MBE growth becomes superconducting below 4 K which contributes a very large diamagnetic signal to the SQUID measurements . \textbf{b,} Remanent magnetization measurements of Mn-Bi$_2$Se$_3$ film with zero field cooling (ZFC). The big drop of magnetization signal comes from the indium layer at the back side of the sample. \textbf{c,} Hysteresis measurements at T=1.5 K. The anomaly at $\textrm{H}\sim0$ again comes from the diamagnetic signal of the superconducting indium layer on the back side of the sample.}
\end{figure*}

As a starting point, the magnetic property of the Mn-Bi$_2$Se$_3$ film can be measured by superconducting quantum interference device (SQUID). As shown by the remanent magnetization measurements on the samples in Fig.~\ref{SQUID}b, clear remanent signal can be observed for temperatures below 10 K. Clear hysteresis is also shown by SQUID below 5 K (Fig.~\ref{SQUID}c). The anomalous drop of the magnetization signal in both remanence and hysteresis comes from the diamagnetic signal of the superconducting indium layer (below 4 K) on the back of the film sample, which is not relevant to the magnetic property of the Mn-Bi$_2$Se$_3$ film itself at all.

In general, SQUID measurements are not so sensitive to small magnetization signals. In fact, small remanent signal is observed by SQUID above 10 K, but the signal is too small and no clear hysteresis is observed as a result. Moreover, SQUID includes magnetization signal of the entire sample (even the irrelevant indium superconducting diamagnetic signal on the back side of the sample). Thus a surface sensitive magnetometer with a better sensitivity to small magnetization signal is needed to measure the magnetic properties on the top surface of the Mn-Bi$_2$Se$_3$ film, where the topological surface electrons are localized, as well as where ARPES and spin-resolved ARPES mainly probe (within $\sim20$ $\textrm{\AA}$).

\textbf{5.2. XMCD measurements on the top surface of the Mn-Bi$_2$Se$_3$ film}

The magnetic properties of the surface of the Mn-Bi$_2$Se$_3$ film need to be treated carefully for two reasons: (i) It has been theoretically predicted that the ferromagnetic order of the TI surface can be achieved via the Ruderman-Kittel-Kasuya-Yosida (RKKY) interaction mediated through the topological surface Dirac fermions \cite{RKYY}, even in absence of the bulk magnetic ordering \cite{Chen Science Fe}. (ii) More importantly, careful Mn concentration characterization measurements have found that the surface has much higher Mn concentration than the entire film crystal \cite{STM}. The Mn concentration of the entire film crystal, which is the nominal concentration used in the maintext, is measured by the in-situ Bi/Mn flux ratio during the MBE growth in conjunction with ex-situ Rutherford back scattering, both of which reveal the ratio of Mn atoms versus the Bi atoms over the entire crystal. On the other hand, the depth profile of Mn doping concentration can be measured by secondary ion mass spectrometry (SIMS), which shows that an excess of Mn accumulates on the surface during the sample growth. SIMS shows that for the top 50 $\textrm{\AA}$ of the surface, the Mn concentration is about 8 times higher than deep in the bulk. Finally, scanning tunneling microscopy (STM), which solely probes the top surface, also reveals a much higher Mn concentration on the surface as compared to the nominal Mn concentration, as well as the the fact that Mn atoms are inhomogeneously distributed on the surface, including the presence of clusters of Mn atoms on the sample surface \cite{STM}. The high Mn concentration on the surface is very likely to be due to the diffusion of Mn atoms toward the film surface at high temperatures ($>200$ K) during MBE growth or during decapping procedure used for ARPES and STM measurements. In the well-known dilute magnetic semiconductors (Ga, Mn)As, the Curie temperature is found to be proportional to the Mn concentration. Mean-field Zener model \cite{GaMnAs} predicts $\textrm{T}_c{\propto}x{\cdot}p^{1/3}$, where $x$ is the doping concentration of Mn to Ga, and $p$ is the hole carrier density. Therefore the high Mn concentration on Bi$_2$Se$_3$ film surface and its p-type doping to the Bi$_2$Se$_3$ system are both favorable to the ferromagnetic order on the surface at the relatively high temperatures($>10$ K).

Here we use the total electron yield (TEY) mode of X-ray magnetic circular dichroism (XMCD, for review of the technique and its application to transition metal thin films and monolayers, see \cite{XMCD_review1, XMCD_review2} ) to directly measure the magnetization of the surface.

\textbf{5.2.1. XMCD on Mn-Bi$_2$Se$_3$ film top surface at T=45 K}
\begin{figure*}[h]
\centering
\begin{center}
\includegraphics[width=18cm]{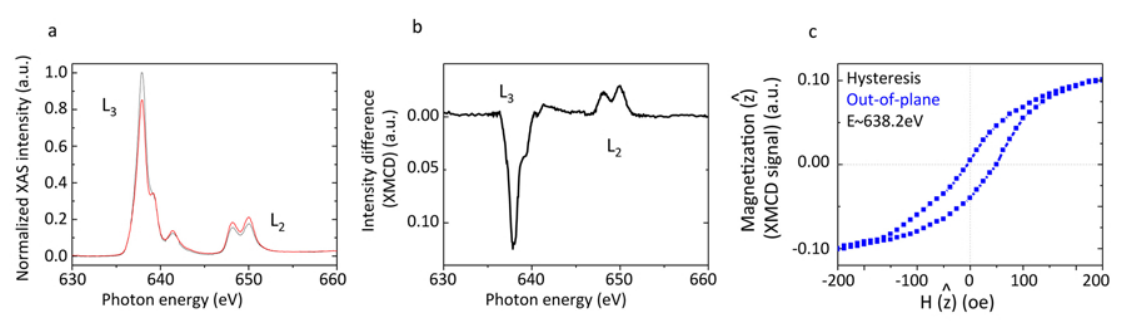}
\end{center}
\caption{\label{XMCD_Mn_45} \textbf{XMCD measurements at Mn $\textrm{L}_{23}$ absorption edge on Mn-Bi$_2$Se$_3$ film top surface at T=45 K.} \textbf{a,} Remanent normalized XAS intensity of Mn $\textrm{L}_{23}$ absorption edge. Black (red) represents the XAS with the photon spin parallel (anti-parallel) with the remanent magnetization. \textbf{b,} The difference of the two XAS (parallel and anti-parallel) generates the XMCD signal, which is proportional to the magnetization of the Mn-Bi$_2$Se$_3$ film top surface. \textbf{c,} Hysteresis measurements are performed with field applied along out-of-plane direction. The measurements are done by measuring XMCD signal at the $\textrm{L}_3$ edge energy $\textrm{E}{\sim}638.2$ eV at different H fields.}
\end{figure*}

Fig.~\ref{XMCD_Mn_45} shows the XMCD measurements on Mn-Bi$_2$Se$_3$ film top surface. The remanent measurements are performed at T=45 K and H=0T after applying magnetic field H=0.5T along -z direction. If the surface is ferromagnetically ordered at 45 K, then a finite remanent magnetization $\vec{\textrm{M}}$ should take place along -z direction after applying the magnetic field. Such remanent magnetization then will lead to a different response of the XAS spectrum using circular plus or circular minus polarized light, resulting in a nonzero remanent XMCD. Indeed, a clear difference between the different photon spin XAS spectra are observed, as shown in Fig.~\ref{XMCD_Mn_45}a. The $\textrm{L}_3$ edge is observed to be enhanced with photon spin parallel to the magnetization, whereas the $\textrm{L}_2$ edge is enhanced with photon spin anti-parallel to the remanent magnetization. The clear remanent XMCD (Fig.~\ref{XMCD_Mn_45}b) suggests that the surface is ordered with easy axis along the out-of-plane direction at T=45 K.

The hysteresis then can be measured by measuring XMCD signal (we choose the peak energy ofs $\textrm{L}_3$ edge $\textrm{E}{\sim}638.2$ eV) at different applied H fields. As shown in Fig.~\ref{XMCD_Mn_45}c, a clear hysteresis is observed at T=45 K along out-of-plane direction, which proves the ferromagnetic order of the Mn-Bi$_2$Se$_3$ film top surface.

\textbf{5.2.2. XMCD temperature dependence}
\begin{figure*}[h]
\centering
\begin{center}
\includegraphics[width=8cm]{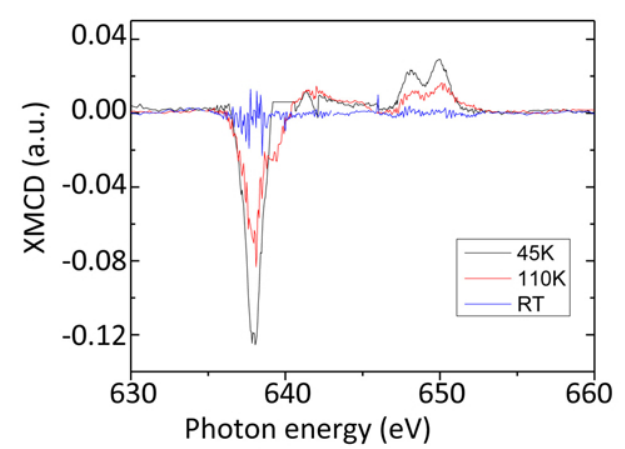}
\end{center}
\caption{\label{XMCD_Mn_tem} \textbf{Temperature dependence of remanent XMCD.} XMCD measurements are performed at three different temperatures: T=45 K, 110 K and 300 K (room temperature).}
\end{figure*}

Fig.~\ref{XMCD_Mn_tem} shows the temperature dependence of the XMCD signal. The XMCD signal is proportional to the remanent magnetization of the film surface, which can be obtained by the sum rules \cite{Sum} (see below). Thus, the temperature dependence of the remanent XMCD signal reveals the temperature dependence of the magnetic ordering of the film surface. As shown in Fig.~\ref{XMCD_Mn_tem}, clear XMCD is observed at T=45 K, whereas the XMCD is strongly suppressed at T=110 K. Finally no XMCD is observed from the measurement performed at room temperature. The detailed Curie temperature of the surface requires future finer measurements with more intermediate temperatures.
\end{enumerate}

\newpage
\textbf{{\large SI IV. Zn-Bi$_2$Se$_3$}}
\begin{enumerate}
\item \textbf{Detailed spin-resolved measurements on Zn-Bi$_2$Se$_3$}

\begin{figure*}[h]
\centering
\begin{center}
\includegraphics[width=15cm]{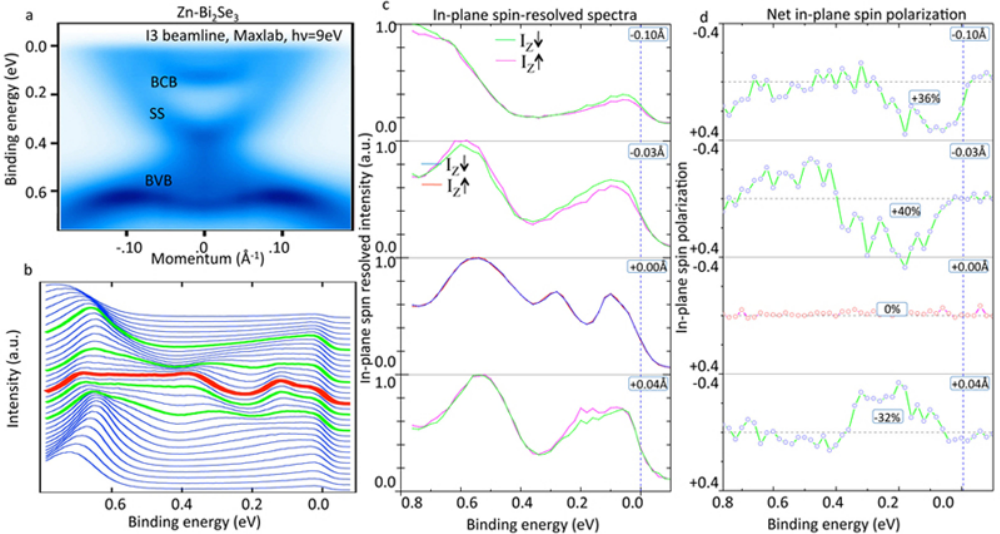}
\end{center}
\caption{\label{Zn_in_plane} \textbf{In-plane spin measurements of Zn-Bi$_2$Se$_3$.} \textbf{a-b,} Spin-integrated ARPES measured dispersion and EDCs with incident photon energy 9 eV (used for spin-resolved measurements). The EDCs selected for spin-resolved measurements are highlighted in green (red) colors in the EDC panel. The EDC at the $\bar{\Gamma}$ momenta is in red color. \textbf{c,} In-plane spin-resolved EDC spectra. \textbf{d,} In-plane spin polarization obtained from \textbf{c}.}
\end{figure*}

\begin{figure*}[h]
\centering
\begin{center}
\includegraphics[width=15cm]{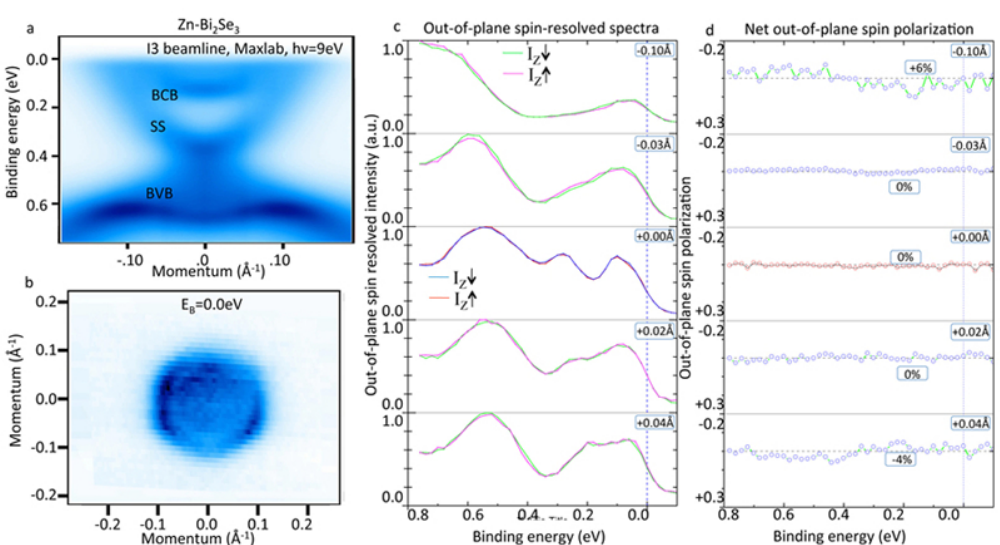}
\end{center}
\caption{\label{Zn_out_of_plane} \textbf{Out-of-plane spin measurements of Zn-Bi$_2$Se$_3$.} \textbf{a,} Spin-integrated ARPES measured dispersion with incident photon energy 9 eV (used for spin-resolved measurements). \textbf{b,} Spin-integrated Fermi surface mapping of Zn-Bi$_2$Se$_3$ at the native chemical potential ($\textrm{E}_\textrm{B}=0.0$ eV) \textbf{c,} Out-of-plane spin-resolved EDC spectra. \textbf{d,} Out-of-plane spin polarization obtained from \textbf{c}.}
\end{figure*}

Here we show comprehensive spin-resolved measurements on Zn-Bi$_2$Se$_3$. These measurements on Zn-Bi$_2$Se$_3$ are performed at the I3 beamline at Maxlab \cite{I3 1,I3 2} in Lund, Sweden, with incident photon energy 9 eV as a control group to the Mn-Bi$_2$Se$_3$. For in-plane spin polarization (see Fig.~\ref{Zn_in_plane}), the spin-helical configuration is again observed without surprise. On the other hand, the out-of-plane spin polarization (see Fig.~\ref{Zn_out_of_plane}) is zero at small momentum near the $\bar{\Gamma}$ momenta which is in sharp contrast to the Mn-Bi$_2$Se$_3$ case. At large momentum ($k\sim-0.10$ $\textrm{\AA}$) a very small but finite $\textrm{P}_z$ appears. However, the $\textrm{P}_z$ in Zn-Bi$_2$Se$_3$ respects TR symmetry since it reverses its sign upon going from $-k$ to $+k$. As shown by Fig.~\ref{Zn_out_of_plane}b, the Fermi surface of Zn-Bi$_2$Se$_3$ is warped into hexagon-shape. Therefore, the observed $\textrm{P}_z$ should be a result of the Fermi surface warping on the Zn-Bi$_2$Se$_3$ Dirac cone \cite{Liang Fu Warping}.
\end{enumerate}

\newpage
\textbf{{\large SI V. Ultra-thin undoped Bi$_2$Se$_3$}}
\begin{enumerate}
\item \textbf{SR-APRES measurements, 3 QL vs 60 QL}
\begin{figure*}[h]
\centering
\begin{center}
\includegraphics[width=17cm]{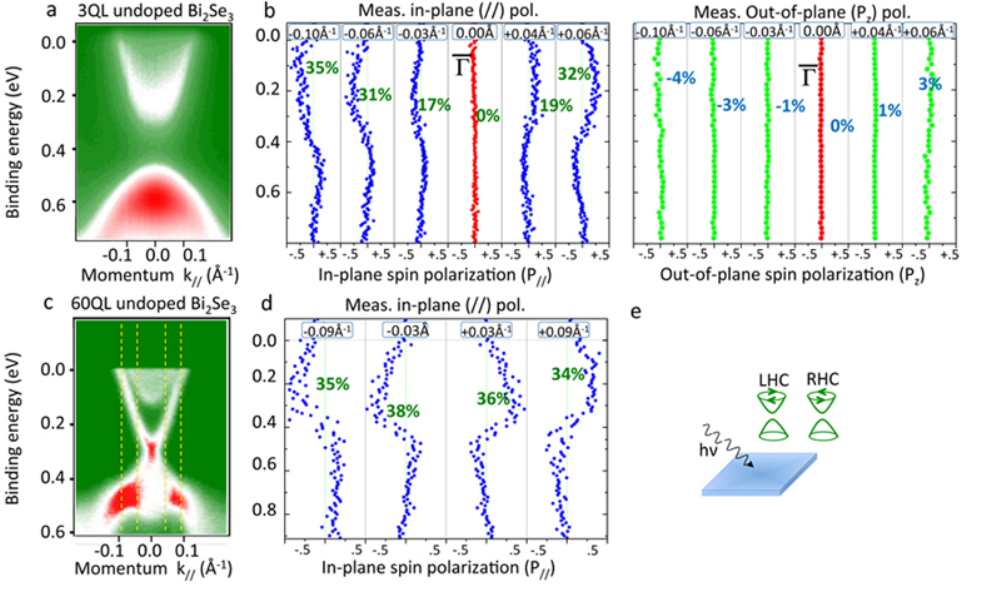}
\end{center}
\caption{\label{QL_Resolution} \textbf{SR-APRES measurements, 3 QL vs 60 QL, demonstrating the momentum resolution of the instrument.} \textbf{a-b,} SR-ARPES measurements ($h\nu=60$ eV) on 3 QL undoped Bi$_2$Se$_3$ shows reduction of spin polarization due at small momenta near the gap. Out-of-plane spin polarization does not show any significant spin polarization $\textrm{P}_z$, especially at the vicinity of the $\bar{\Gamma}$ momenta. \textbf{c-d,} SR-ARPES measurements on 60 QL undoped Bi$_2$Se$_3$ measured with identical experimental settings ($h\nu=60$ eV) does not show any reduction of spin polarization. \textbf{e,} in 3 QL ultra-thin film, the surfaces from top and bottom are quantum mechanically coupled to each other, resulting in a tunneling gap as observed in panel \textbf{a}.}
\end{figure*}

Here we show the spin polarization measurements on 3 QL undoped Bi$_2$Se$_3$ (Fig. 4e-h in the maintext) again, and comparing the results with the spin polarization measurements on 60 QL undoped Bi$_2$Se$_3$. The thickness of 60 QL is well-above the quantum tunneling thickness limit. Thus no tunneling or coupling effect between the top and bottom surfaces is expected. This can also be seen from the ARPES measured \textit{gapless} dispersion, as shown in Fig.~\ref{QL_Resolution}c. The 3 QL and 60 QL SR measurements are performed with identical experimental conditions, except the film thickness. As shown in Fig.~\ref{QL_Resolution}d, no reduction of the spin polarization is observed on 60 QL film surface, even at small momentum such as $\pm0.03$ $\textrm{\AA}^{-1}$. Such measurements demonstrate the momentum resolution of the instruments, and show that the reduction of spin polarization on 3 QL film surface is not caused by blurring between the two branches of the upper Dirac cone due to the momentum resolution of the instrument.
\end{enumerate}

\end{document}